\documentclass{article}%
\usepackage{graphicx}
\usepackage{amsmath}
\usepackage{amsfonts}
\usepackage{amssymb}%
\newtheorem{theorem}{Theorem}
{}

\newtheorem{definition}{Definition}

\newtheorem{lemma}{Lemma}
{}

\newtheorem{remark}{Remark}

\newenvironment{proof}[1][Proof]{\textbf{#1.} }{\ \rule{0.5em}{0.5em}}

\begin{document}

\author{O. A. Veliev\\{\small \ Dept. of Math, Fen-Ed. Fak, Dogus University.,}\\{\small Acibadem, Kadikoy, Istanbul, Turkey,}\\{\small \ e-mail: oveliev@dogus.edu.tr}}
\title{\textbf{On the Polyharmonic Operator with a Periodic Potential}}
\date{}
\maketitle

\begin{abstract}
In this paper we obtain the asymptotic formulas of arbitrary order for the
Bloch eigenvalues and Bloch functions of the $d$-dimensional polyharmonic
operator $L(l,q(x))=(-\Delta)^{l}+q(x)$ with periodic, with respect to
\ arbitrary lattice, potential $q(x),$ where $l\geq1$ and $d\geq2.$ Then we
prove that the number of gaps in the spectrum of the operator $L(l,q(x))$ is
finite. In particular, taking $l=1,$ we get the proof of the Bethe -Sommerfeld
conjecture for arbitrary dimension and arbitrary lattice.

\end{abstract}

\section{Introduction}

\bigskip In this paper we consider the operator%
\begin{equation}
L(l,q(x))=(-\Delta)^{l}+q(x),\ x\in\mathbb{R}^{d},\ d\geq2,l\geq1
\end{equation}
with a periodic (relative to a lattice $\Omega$) potential $q(x)\in W_{2}%
^{s}(F),$ where

$s\geq s_{0}=\frac{3d-1}{2}(3^{d}+d+2)+\frac{1}{4}d3^{d}+d+6,$ $F\equiv
\mathbb{R}^{d}/\Omega$ is a fundamental domain of $\Omega.$ Without loss of
generality it can be assumed that the measure $\mu(F)$ of $F$ is $1$ and
$\int_{F}q(x)dx=0.$ Let $L_{t}(l,q(x))$ be the operator generated in $F$ by
(1) and the conditions:%
\begin{equation}
u(x+\omega)=e^{i(t,\omega)}u(x),\ \forall\omega\in\Omega,
\end{equation}
where $t\in F^{\star}\equiv\mathbb{R}^{d}/\Gamma$ and $\Gamma$ is the lattice
dual to $\Omega$, that is, $\Gamma$ is the set of all vectors $\gamma
\in\mathbb{R}^{d}$ satisfying $(\gamma,\omega)\in2\pi Z$ for all $\omega
\in\Omega.$ It is well-known that the spectrum of the operator $L_{t}(l,q(x))$
consists of the eigenvalues

$\Lambda_{1}(t)\leq\Lambda_{2}(t)\leq....$The function $\Lambda_{n}(t)$ is
called $n$-th band function and its range $A_{n}=\left\{  \Lambda_{n}(t):t\in
F^{\ast}\right\}  $ is called the $n$-th band of the spectrum $Spec(L)$ of $L$
and $Spec(L)=\cup_{n=1}^{\infty}A_{n}$. The eigenfunction $\Psi_{n,t}(x)$ of
$L_{t}(l,q(x))$ corresponding to the eigenvalue $\Lambda_{n}(t)$ is known as
Bloch functions. In the case $q(x)=0$ these eigenvalues and eigenfunctions are
$\mid\gamma+t\mid^{2l}$ and $e^{i(\gamma+t,x)}$ for $\gamma\in\Gamma$.

This paper consists of 4 section. First section is the introduction, where we
describe briefly the scheme of this paper and discuss the related papers.

Let the potential $q(x)$ be a trigonometric polynomial%

\[
\sum_{\gamma\in Q}q_{\gamma}e^{i(\gamma,x)},
\]
where $q_{\gamma}=(q(x),e^{i(\gamma,x)})=\int_{F}q(x)e^{-i(\gamma_{1},x)}dx,$
and $Q=\{\gamma\in\Gamma:q_{\gamma}\neq0\}$ consists of a finite number of
vectors $\gamma$ from $\Gamma.$ Then the eigenvalue $\left\vert \gamma
+t\right\vert ^{2l}$ \ is called a non-resonance eigenvalue if $\gamma+t$ does
not belong to any of the sets

$W_{b,\alpha_{1}}^{l}=\{x\in\mathbb{R}^{d}:\mid\mid x\mid^{2l}-\mid
x+b\mid^{2l}\mid<\mid x\mid^{\alpha_{1}}\},$ that is, if $\gamma+t$ lies far
from the diffraction hyperplanes $D_{b}=\{x\in\mathbb{R}^{d}:\mid x\mid
^{2}=\mid x+b\mid^{2}\},$ where $\alpha_{1}\in(0,1),$ $b\in Q$ . The idea of
the definition of the non-resonance eigenvalue $\left\vert \gamma+t\right\vert
^{2l}$ \ is the following. If \ $\gamma+t\notin W_{b,\alpha_{1}}^{l}$ then the
influence of $q_{b}e^{i(b,x)}$ to the eigenvalue $\left\vert \gamma
+t\right\vert ^{2l}$ is not significant. If $\gamma+t$ does not belong to any
of the sets $W_{b,\alpha_{1}}^{l}$ for $b\in Q$ then the influence of the
trigonometric polynomial $q(x)$ to the eigenvalue $\left\vert \gamma
+t\right\vert ^{2l}$ is not significant. Therefore the corresponding
eigenvalue of the operator $L_{t}(l,q(x))$ is close to the eigenvalue
$\mid\gamma+t\mid^{2l}$ of $L_{t}(l,0).$

If $q(x)\in W_{2}^{s}(F),$ then to describe the non-resonance and resonance
eigenvalues $\left\vert \gamma+t\right\vert ^{2l}$ of the order of $\rho^{2l}$
( written as $\left\vert \gamma+t\right\vert \sim\rho$) for big parameter
$\rho$ we write the potential $q(x)\in W_{2}^{s}(F)$ in the form
\begin{equation}
q(x)=\sum_{\gamma_{1}\in\Gamma(\rho^{\alpha})}q_{\gamma_{1}}e^{i(\gamma
_{1},x)}+O(\rho^{-p\alpha}),
\end{equation}
where $\Gamma(\rho^{\alpha})=\{\gamma\in\Gamma:0<$ $\mid\gamma\mid
<\rho^{\alpha})\}$, $p=s-d,$ $\alpha=\frac{1}{m},$ $m=3^{d}+d+2,$ and the
relation $\left\vert \gamma+t\right\vert \sim\rho$ means that $c_{1}%
\rho<\left\vert \gamma+t\right\vert <c_{2}\rho$. Here and in subsequent
relations we denote by $c_{i}$ ($i=1,2,...)$ the positive constants,
independent on $\rho,$ whose exact values are inessential. Note that $q(x)\in
W_{2}^{s}(F)$ means that $\sum_{\gamma}\mid q_{\gamma}\mid^{2}(1+\mid
\gamma\mid^{2s})<\infty.$ If $s\geq d,$ then
\begin{equation}
\sum_{\gamma}\mid q_{\gamma}\mid<c_{3},\text{ }\sup_{x\in\lbrack0,1]}\mid
\sum_{\gamma\notin\Gamma(c_{1}\rho^{\alpha})}q_{\gamma}e^{i(\gamma,x)}\mid
\leq\sum_{\mid\gamma\mid\geq c_{1}\rho^{\alpha}}\mid q_{\gamma}\mid
=O(\rho^{-p\alpha}),
\end{equation}
i.e., (3) holds. It follows from (4) that the influence of $\sum_{\gamma
\notin\Gamma(c_{1}\rho^{\alpha})}q_{\gamma}e^{i(\gamma,x)}$ to the eigenvalue
$\left\vert \gamma+t\right\vert ^{2l}$ is $O(\rho^{-p\alpha}).$ If $\gamma+t$
does not belong to any of the sets

$W_{b,\alpha_{1}}^{l}(c_{2})=\{x\in\mathbb{R}^{d}:\mid\mid x\mid^{2l}-\mid
x+b\mid^{2l}\mid<c_{2}\mid x\mid^{\alpha_{1}}\}$ for $b\in\Gamma(c_{1}%
\rho^{\alpha}),$ then the influence of the trigonometric polynomial
$P(x)=\sum_{\gamma\in\Gamma(c_{1}\rho^{\alpha})}q_{\gamma}e^{i(\gamma,x)}$ to
the eigenvalue $\left\vert \gamma+t\right\vert ^{2l}$ is not significant. Thus
the corresponding eigenvalue of the operator $L_{t}(l,q(x))$ is close to the
eigenvalue $\mid\gamma+t\mid^{2l}$ of $L_{t}(l,0).$ Note that changing the
values of $c_{1}$ and $c_{2}$\ in the definitions of $W_{b,\alpha_{1}}%
^{l}(c_{2})$ and $P(x)$ we obtain the different definitions of the
non-resonance eigenvalues. However, in any case we obtain the same asymptotic
formulas and the same perturbation theory, that is, this changing does not
change anything for asymptotic formulas. Therefore we can define the
non-resonance eigenvalue in different way. In accordance with the case of the
trigonometric polynomial it is natural to say that the eigenvalue $\left\vert
\gamma+t\right\vert ^{2l}$ is a non-resonance eigenvalue if $\gamma+t$ does
not belong to any of the sets $W_{b,\alpha_{1}}^{l}(c_{2})$ for $\mid
b\mid<c_{1}p\left\vert \gamma+t\right\vert ^{\alpha}$. However, for
simplicity, we give the definitions as follows. By definition, put $\alpha
_{k}=3^{k}\alpha$ for $k=1,2,...$ and introduce the sets

$V_{\gamma_{1}}^{l}(\rho^{\alpha_{1}})\equiv\{x\in\mathbb{R}^{d}:\mid\mid
x\mid^{2l}-\mid x+\gamma_{1}\mid^{2l}\mid<\rho^{\alpha_{1}}\}\cap(R(\frac
{3}{2}\rho)\backslash R(\frac{1}{2}\rho))$%
\[
E_{1}^{l}(\rho^{\alpha_{1}},p)\equiv\bigcup_{\gamma_{1}\in\Gamma(p\rho
^{\alpha})}V_{\gamma_{1}}^{l}(\rho^{\alpha_{1}}),\text{ }U^{l}(\rho
^{\alpha_{1}},p)\equiv(R(\frac{3}{2}\rho)\backslash R(\frac{1}{2}%
\rho))\backslash E_{1}^{l}(\rho^{\alpha_{1}},p),
\]%
\[
E_{k}^{l}(\rho^{\alpha_{k}},p)\equiv\bigcup_{\gamma_{1},\gamma_{2}%
,...,\gamma_{k}\in\Gamma(p\rho^{\alpha})}(\cap_{i=1}^{k}V_{\gamma_{i}}%
^{l}(\rho^{\alpha_{k}})),
\]
where $R(\rho)=\{x\in\mathbb{R}^{d}:\mid x\mid<\rho\},$ $\rho$ is a big
parameter and the intersection $\cap_{i=1}^{k}V_{\gamma_{i}}^{l}$ in the
definition of $E_{k}^{l}$ is taken over $\gamma_{1},\gamma_{2},...,\gamma
_{k},$ that are linearly independent. The set $U^{l}(\rho^{\alpha_{1}},p)$ is
said to be a non-resonance domain and the eigenvalue $\left\vert
\gamma+t\right\vert ^{2l}$ is called a non-resonance eigenvalue if
$\gamma+t\in U^{l}(\rho^{\alpha_{1}},p).$ The domains $V_{\gamma_{1}}^{l}%
(\rho^{\alpha_{1}})$ for $\gamma_{1}\in\Gamma(p\rho^{\alpha})$ are called
resonance domains and $\mid\gamma+t\mid^{2l}$ is called a resonance eigenvalue
if $\gamma+t\in V_{\gamma_{1}}^{l}(\rho^{\alpha_{1}}).$ In Remark 1 we will
discuss the relations between sets $W_{b,\alpha_{1}}^{l}(c_{2})$ and
$V_{b}^{l}(\rho^{\alpha_{1}}).$

In section 2 we prove that for each $\gamma+t\in U^{l}(\rho^{\alpha_{1}},p)$
there exists an eigenvalue $\Lambda_{N}(t)$ of the operator $L_{t}(l,q(x))$
satisfying the following formulae
\begin{equation}
\Lambda_{N}(t)=\mid\gamma+t\mid^{2l}+F_{k-1}(\gamma+t)+O(\mid\gamma
+t\mid^{-3k\alpha})
\end{equation}
for $k=1,2,...,[\frac{1}{3}(p-\frac{1}{2}m(d-1))],$ where $[a]$ denotes the
integer part of $a,$ $F_{0}=0,$ and $F_{k-1}$ ( for $k>1)$ is explicitly
expressed by the potential $q(x)$ and eigenvalues of $L_{t}(0).$ Besides, we
prove that if the conditions
\begin{align}
&  \mid\Lambda_{N}(t)-\mid\gamma+t\mid^{2l}\mid<\frac{1}{2}\rho^{\alpha_{1}%
},\\
&  \mid b(N,\gamma)\mid>c_{4}\rho^{-c\alpha}%
\end{align}
hold, \ where $c$ is a positive constant,
\begin{equation}
b(N,\gamma)=(\Psi_{N,t},e^{i(\gamma+t,x)}),
\end{equation}
$\Psi_{N,t}(x)$ is a normalized eigenfunction of $L_{t}(l,q(x))$ corresponding
to $\Lambda_{N}(t),$ then $\ $the following statements are valid:

(a) if $\gamma+t$ is in the non-resonance domain, then $\Lambda_{N}(t)$
satisfies (5) for $k=1,2,...,[\frac{1}{3}(p-c)]$ ( see Theorem 1);

(b) if $\gamma+t\in E_{s}^{l}\backslash E_{s+1}^{l},$ where $s=1,2,...,d-1,$
then%
\begin{equation}
\Lambda_{N}(t)=\lambda_{j}(\gamma+t)+O(\mid\gamma+t\mid^{-k\alpha}),
\end{equation}
where $\lambda_{j}$ is an eigenvalue of a matrix $C(\gamma+t)$ ( see (27) and
Theorem 2). Moreover, we prove that every big eigenvalue of the operator
$L_{t}(l,q(x))$ for all values of $t$ satisfies one of these formulae.

For investigation of the Bloch function in the non-resonance domain, in
section 3, we find the values of quasimomenta $\gamma+t$ for which the
corresponding eigenvalues are simple , namely we construct the subset $B$ of
$U^{1}(\rho^{\alpha_{1}},p)$ with the following properties:

Pr.1. If $\gamma+t\in B,$ then there exists a unique eigenvalue, denoted by

$\Lambda(\gamma+t),$ of the operator $L_{t}(l,q(x))$ satisfying (5). This is a
simple eigenvalue of $L_{t}(l,q(x))$. Therefore we call the set $B$ the simple
set of quasimomenta.

Pr.2. The eigenfunction $\Psi_{N(\gamma+t)}(x)\equiv\Psi_{\gamma+t}(x)$
corresponding to the eigenvalue $\Lambda(\gamma+t)$ is close to $e^{i(\gamma
+t,x)}$, namely
\begin{equation}
\Psi_{N}(x)=e^{i(\gamma+t,x)}+O(\mid\gamma+t\mid^{-\alpha_{1}}),
\end{equation}%
\begin{equation}
\Psi_{\gamma+t}(x)=e^{i(\gamma+t,x)}+\Phi_{k-1}(x)+O(\mid\gamma+t\mid
^{-k\alpha_{1}}),\text{ }k=1,2,...\text{ ,}%
\end{equation}
where $\Phi_{k-1}$ is explicitly expressed by $q(x)$ and the eigenvalues of
$L_{t}(l,0).$

Pr.3. The set $B$ has asymptotically full measure on $\mathbb{R}^{d}$ and
contains the intervals $\{a+sb:s\in\lbrack-1,1]\}$ such that $\Lambda
(a-b)<\rho^{2l},$ $\Lambda(a+b)>\rho^{2l},$\ and $\Lambda(\gamma+t)$ \ is
continuous on these intervals. Hence there exists $\gamma+t$ such that
$\Lambda(\gamma+t)=\rho^{2l}.$ It implies the validity of Bethe-Sommerfeld
conjecture for $L(l,q(x)).$ These results is proved in section 4.

Construction of the set $B$ consists of two steps.

Step 1. We prove that all eigenvalues $\Lambda_{N}(t)\sim\rho^{2l}$ of the
operator $L_{t}(l,q(x))$ lie in the $\varepsilon_{1}=\rho^{-d-2\alpha}$
neighborhood of the numbers

$F(\gamma+t)=\mid\gamma+t\mid^{2l}+F_{k_{1}-1}(\gamma+t)$, $\lambda_{j}%
(\gamma+t)$ ( see (5), (9)), where $k_{1}=[\frac{d}{3\alpha}]+2.$ We call
these numbers as the known parts of the eigenvalues. Moreover, for
$\gamma+t\in U^{l}(\rho^{\alpha_{1}},p)$ there exists $\Lambda_{N}(t)$
satisfying $\Lambda_{N}(t)=F(\gamma+t)+o(\varepsilon_{1})$.

Step 2. By eliminating the set of quasimomenta $\gamma+t$, for which the known
parts $F(\gamma+t)$ of $\Lambda_{N}(t)$ are situated from the known parts
$F(\gamma^{^{\prime}}+t),$ $\lambda_{j}(\gamma^{^{\prime}}+t)$ ($\gamma
^{^{\prime}}\neq\gamma)$ of other eigenvalues at a distance less than
$2\varepsilon_{1},$ we construct the set $B$ with the following properties: if
$\gamma+t\in B,$ then the following conditions (called simplicity conditions
for $\Lambda_{N}(t))$ hold
\begin{equation}
\mid F(\gamma+t)-F(\gamma^{^{\prime}}+t)\mid\geq2\varepsilon_{1}\text{ }%
\end{equation}
for $\gamma^{^{\prime}}\in K\backslash\{\gamma\},$ $\gamma^{^{\prime}}+t\in
U^{1}(\rho^{\alpha_{1}},p)$ and%
\begin{equation}
\mid F(\gamma+t)-\lambda_{j}(\gamma^{^{\prime}}+t)\mid\geq2\varepsilon_{1}%
\end{equation}
for $\gamma^{^{\prime}}\in K,\gamma^{^{\prime}}+t\in E_{k}^{1}\backslash
E_{k+1}^{1},$ $j=1,2,...,$ where $K$ is the set of $\gamma^{^{\prime}}%
\in\Gamma$ satisfying $\mid F(\gamma+t)-\mid\gamma^{^{\prime}}+t\mid^{2l}%
\mid<\frac{1}{3}\rho^{\alpha_{1}}$. Thus we define the simple set $B$ as follows

\begin{definition}
The simple set $B$ is the set of

$x\in U^{1}(\rho^{\alpha_{1}},p)\cap(R(\frac{3}{2}\rho-\rho^{\alpha_{1}%
-1})\backslash R(\frac{1}{2}\rho+\rho^{\alpha_{1}-1}))$ such that
$x=\gamma+t,$ where $\gamma\in\Gamma,t\in F^{\star},$ and the simplicity
conditions (12), (13) hold.
\end{definition}

As a consequence of these conditions the eigenvalue $\Lambda_{N}(t)$ does not
coincide with other eigenvalues. To prove this, namely to prove the Pr.1 and
(10), we show that for any normalized eigenfunction $\Psi_{N}(x)$
corresponding to $\Lambda_{N}(t)$ the following equality holds:
\begin{equation}
\sum_{\gamma^{^{\prime}}\in\Gamma\backslash\gamma}\mid b(N,\gamma^{^{\prime}%
})\mid^{2}=O(\rho^{-2\alpha_{1}}).
\end{equation}

The listed all results ( division the eigenvalues $\left\vert \gamma
+t\right\vert ^{2l}$, for big $\ \gamma\in\Gamma,$ into two groups:
non-resonance ones and resonance ones, the proof of the formulas (5), (9),
construction and investigations of the simply set $B,$ the proof of the
asymptotic formulas (11) for Bloch function and implication the proof of the
Bethe-Sommerfeld conjecture for arbitrary dimension and arbitrary lattices
from these formulas )\ for the first time were obtained in papers [12-14,16]
for the Schrodinger operator $L(1,q(x)).$ For the first time in [12-14] we
constructed the simple set $B$ with the Pr.1 and Pr.3., though in those papers
we emphasized the Bethe-Zommerfeld conjecture. Note that for this conjecture
and for Pr.1, Pr.3. it is enough to prove that the left-hand side of (14) is
less than $\frac{1}{4}$ ( we proved this inequality in [12-14] and as noted in
Theorem 3 of [13] and in [16] the proof of this inequality does not differ
from the proof of (14)). From (10) we got \ (11) by iteration (see [16]) . The
enlarged form of this results is written in [15],[18],[19].

The main difficulty and the crucial point of papers [12-14] were the
construction and investigations of the simple set $B$ of quasimomenta in
neighborhood of the surface $\{\gamma+t\in U^{1}(\rho^{\alpha_{1}}%
,p):F(\gamma+t)=\rho^{2}\}.$ This difficulty of the perturbation theory of
$L(1,q(x))$ is of a physical nature and it is connected with the complicated
picture of the crystal diffraction. If $d=2,3,$ then $F(\gamma+t)=\mid
\gamma+t\mid^{2}$ and the matrix $C(\gamma+t)$ corresponds to the Schrodinger
operator with directional potential $q_{\gamma_{1}}(x)=\sum_{n\in Z}%
q_{n\gamma_{1}}e^{i(n\gamma_{1},x)}$ ( see [13]). So for construction of the
simple set $B$ of quasimomenta we eliminated the vicinities of the diffraction
planes and the sets connected with directional potential ( see (12), (13)).
Besides,\ for nonsmooth potentials $q(x)\in L_{2}(\mathbb{R}^{2}/\Omega),$we
eliminated a set, which is described in the terms of the number of states (
see [12,17]). The simple sets $B$ of quasimomenta for the first time are
constructed and investigated ( hence also the main difficulty and the crucial
point of perturbation theory of $L(1,q)$ is investigated) in [13] for $d=3$
and in [12,14] for the cases:

1. $d=2,$ $q(x)\in L_{2}(F)$ ;

2. $d>2,$ $q(x)$ is a smooth potential.

Then, Yu.E. Karpeshina proved ( see [6],[7],[8]) the convergence of the
perturbation series of $L(l,q)$ with a wide class of nonsmooth potentials
$q(x)$ for a set, that is similar to $B$, of quasimomenta in the cases:

1. $2l>d;$ 2. $4l>d+1,$ $(2l\leq d)$; 3. $d=3,l=1,$ and using it she proved
the validity of \ the Bethe-Sommerfeld conjecture in these cases. In papers
[2,3] asymptotic formulas for eigenvalues and Bloch function of two and three
dimensional operator $L_{t}(1,q(x))$ were obtained. In [4] asymptotic formulae
for non-resonance eigenvalues of $L_{0}(1,q(x))$ were obtained.

For the first time M.M. Skriganov [10,11] proved the validity of the
Bethe-Sommerfeld conjecture for the Schrodinger operator for dimension $d=2,3$
for arbitrary lattice, for dimension $d>3$ for rational lattice, and for the
operator $L(l,q(x))$ for $2l>d.$ The Skriganov's method is based on the detail
investigation of the arithmetic and geometric properties of the lattice.
B.E.J.Dahlberg \ and E.Trubowits [1] using an asymptotic of Bessel function,
gave the beautiful proof of this conjecture for the two dimensional Scrodinger
operator. B. Helffer and A. Mohamed [5], by investigations the integrated
density of states, proved the validity of the Bethe-Sommerfeld conjecture for
the Scrodinger operator for $d\leq4,$ for arbitrary lattice. Recently
Parnovski and Sobelev [9] proved this conjecture for the operator $L(l,q(x)),$
for $8l>d+3.$

The method of this paper and papers [12-14] is a first and unique, for the
present, by which the validity of the Bethe-Sommerfeld conjecture for
arbitrary lattice and for arbitrary dimension is proved. For the operator
$L(l,q),$ in order to avoid eclipsing the essence by technical details,\ we
assume that $l\geq1.$ It can be replaced by $l>n_{s,d}$ , where $n_{s,d}<1$
and depends on the smoothness $s$ of the potential $q(x)\in W_{2}%
^{s}(\mathbb{R}^{d}/\Omega)$ and the dimension $d$ .

In this paper for the different types of the measures of the subset $A$ of
$\mathbb{R}^{d}$ we use the same notation $\mu(A).$ By $\mid A\mid$ we denote
the number of elements of the set $A\subset\Gamma$ and use the following
obvious fact. If $a\sim\rho,$ then the number of elements of the set
$\{\gamma+t:$ $\gamma\in\Gamma\}$ satisfying $\mid\mid\gamma+t\mid-a\mid<1$ is
less than $c_{5}\rho^{d-1}.$ Therefore the number of eigenvalues of
$L_{t}(l,q)$ lying in $(a^{2l}-\rho^{2l-1},a^{2l}+\rho^{2l-1})$ is less than
$c_{5}\rho^{d-1}.$ Besides, we use the inequalities:%
\begin{align}
\alpha_{1}+d\alpha &  <1-\alpha\,,\ \ \ \ \ d\alpha<\frac{1}{2}\alpha
_{d},\ \ \ k_{1}\leq\frac{1}{3}(p-\frac{1}{2}(m(d-1)),\\
p_{1}\alpha_{1}  &  \geq p\alpha,\ \ \ \ \ 3k_{1}\alpha>d+2\alpha
,\ \ \ \ \ \ \alpha_{k}+(k-1)\alpha<1,\nonumber\\
\alpha_{k+1}  &  >2(\alpha_{k}+(k-1))\alpha\nonumber
\end{align}
for $k=1,2,...,d,$ which follow from the definitions $p=s-d,$ $\alpha
_{k}=3^{k}\alpha,$ $\alpha=\frac{1}{m},$ $m=3^{d}+d+2,$ $k_{1}=[\frac
{d}{3\alpha}]+2,$ $p_{1}=[\frac{p}{3}]+1$ of the numbers $p,m,\alpha
_{k},\alpha,k_{1},p_{1}.$

\section{Asymptotic Formulae for Eigenvalues}

In this section we obtain the asymptotic formulas for the eigenvalues by
iteration of the formula
\begin{equation}
(\Lambda_{N}-\mid\gamma+t\mid^{2l})b(N,\gamma)=(\Psi_{N,t}(x)q(x),e^{i(\gamma
+t,x)}),
\end{equation}
where $\gamma+t\in U^{l}(\rho^{\alpha_{1}},p)$ and $b(N,\gamma)$ is defined in
(8). Introducing into (16) the expansion (3) of $q(x)$, we get
\begin{equation}
(\Lambda_{N}-\mid\gamma+t\mid^{2l})b(N,\gamma)=\sum_{\gamma_{1}\in\Gamma
(\rho^{\alpha})}q_{\gamma_{1}}b(N,\gamma-\gamma_{1})+O(\rho^{-p\alpha}).
\end{equation}
From the relations (16), (17) it follows that
\begin{equation}
b(N,\gamma^{^{\prime}})=\dfrac{(\Psi_{N,t}q(x),e^{i(\gamma^{^{\prime}}+t,x)}%
)}{\Lambda_{N}-\mid\gamma^{^{\prime}}+t\mid^{2l}}=%
%TCIMACRO{\dsum _{\gamma_{1}\in\Gamma(\rho^{\alpha})}}%
%BeginExpansion
{\displaystyle\sum_{\gamma_{1}\in\Gamma(\rho^{\alpha})}}
%EndExpansion
\dfrac{q_{\gamma_{1}}b(N,\gamma^{^{\prime}}-\gamma_{1})}{\Lambda_{N}%
-\mid\gamma^{^{\prime}}+t\mid^{2l}}+O(\rho^{-p\alpha})
\end{equation}
for all vectors $\gamma^{^{\prime}}\in\Gamma$ satisfying the inequality
\begin{equation}
\mid\Lambda_{N}-\mid\gamma^{^{\prime}}+t\mid^{2l}\mid>\frac{1}{2}\rho
^{\alpha_{1}}.
\end{equation}
If (6) holds and $\gamma+t\in U^{l}(\rho^{\alpha_{1}},p),$ then
\begin{equation}
\mid\mid\gamma+t\mid^{2l}-\mid\gamma-\gamma_{1}+t\mid^{2l}\mid>\rho
^{\alpha_{1}},\text{ }\mid\Lambda_{N}-\mid\gamma-\gamma_{1}+t\mid^{2l}%
\mid>\frac{1}{2}\rho^{\alpha_{1}}\text{ }%
\end{equation}
for all $\gamma_{1}\in\Gamma(p\rho^{\alpha}).$ Hence the vector $\gamma
-\gamma_{1}$ for $\gamma+t\in U(\rho^{\alpha_{1}},p)$ and $\gamma_{1}\in
\Gamma(p\rho^{\alpha})$ satisfies (19). Therefore, in (18) one can replace
$\gamma^{^{\prime}}$ by $\gamma-\gamma_{1}$ and write
\[
b(N,\gamma-\gamma_{1})=%
%TCIMACRO{\dsum _{\gamma_{2}\in\Gamma(\rho^{\alpha})}}%
%BeginExpansion
{\displaystyle\sum_{\gamma_{2}\in\Gamma(\rho^{\alpha})}}
%EndExpansion
\dfrac{q_{\gamma_{2}}b(N,\gamma-\gamma_{1}-\gamma_{2})}{\Lambda_{N}-\mid
\gamma-\gamma_{1}+t\mid^{2l}}+O(\rho^{-p\alpha}).
\]
Substituting this for $b(N,\gamma-\gamma_{1})$ into the right-hand side of
(17) and isolating the terms containing the multiplicand $b(N,\gamma)$, we
get
\[
(\Lambda_{N}-\mid\gamma+t\mid^{2l})b(N,\gamma)=%
%TCIMACRO{\dsum _{\gamma_{1},\gamma_{2}\in\Gamma(\rho^{\alpha})}}%
%BeginExpansion
{\displaystyle\sum_{\gamma_{1},\gamma_{2}\in\Gamma(\rho^{\alpha})}}
%EndExpansion
\dfrac{q_{\gamma_{1}}q_{\gamma_{2}}b(N,\gamma-\gamma_{1}-\gamma_{2})}%
{\Lambda_{N}-\mid\gamma-\gamma_{1}+t\mid^{2l}}+O(\rho^{-p\alpha})=
\]%
\[%
%TCIMACRO{\dsum _{\gamma_{1}\in\Gamma(\rho^{\alpha})}}%
%BeginExpansion
{\displaystyle\sum_{\gamma_{1}\in\Gamma(\rho^{\alpha})}}
%EndExpansion
\dfrac{\mid q_{\gamma_{1}}\mid^{2}b(N,\gamma)}{\Lambda_{N}-\mid\gamma
-\gamma_{1}+t\mid^{2l}}+%
%TCIMACRO{\dsum _{\substack{\gamma_{1},\gamma_{2}\in\Gamma(\rho^{\alpha
%}),\\\gamma_{1}+\gamma_{2}\neq0}}}%
%BeginExpansion
{\displaystyle\sum_{\substack{\gamma_{1},\gamma_{2}\in\Gamma(\rho^{\alpha
}),\\\gamma_{1}+\gamma_{2}\neq0}}}
%EndExpansion
\dfrac{q_{\gamma_{1}}q_{\gamma_{2}}b(N,\gamma-\gamma_{1}-\gamma_{2})}%
{\Lambda_{N}-\mid\gamma-\gamma_{1}+t\mid^{2l}}+O(\rho^{-p\alpha}),
\]
since $q_{\gamma_{1}}q_{\gamma_{2}}=\mid q_{\gamma_{1}}\mid^{2}$ for
$\gamma_{1}+\gamma_{2}=0$ and the last summation is taken under the condition
$\gamma_{1}+\gamma_{2}\neq0.$ Repeating this process $p_{1}\equiv\lbrack
\frac{p}{3}]+1$ times, i.e., in the last summation replacing $b(N,\gamma
-\gamma_{1}-\gamma_{2})$ by its expression from (18) ( in (18) replace
$\gamma^{^{\prime}}$ by $\gamma-\gamma_{1}-\gamma_{2}$) and isolating the
terms containing $b(N,\gamma)$ etc., we obtain
\begin{equation}
(\Lambda_{N}-\mid\gamma+t\mid^{2l})b(N,\gamma)=A_{p_{1}}(\Lambda_{N}%
,\gamma+t)b(N,\gamma)+C_{p_{1}}+O(\rho^{-p\alpha}),
\end{equation}
where $A_{p_{1}}(\Lambda_{N},\gamma+t)=\sum_{k=1}^{p_{1}}S_{k}(\Lambda
_{N},\gamma+t)$ ,
\[
S_{k}(\Lambda_{N},\gamma+t)=%
%TCIMACRO{\dsum _{\gamma_{1},...,\gamma_{k}\in\Gamma(\rho^{\alpha})}}%
%BeginExpansion
{\displaystyle\sum_{\gamma_{1},...,\gamma_{k}\in\Gamma(\rho^{\alpha})}}
%EndExpansion
\dfrac{q_{\gamma_{1}}q_{\gamma_{2}}...q_{\gamma_{k}}q_{-\gamma_{1}-\gamma
_{2}-...-\gamma_{k}}}{\prod_{j=1}^{k}(\Lambda_{N}-\mid\gamma+t-\sum_{i=1}%
^{j}\gamma_{i}\mid^{2l})},
\]%
\[
C_{p_{1}}=\sum_{\gamma_{1},...,\gamma_{p_{1}+1}\in\Gamma(\rho^{\alpha})}%
\dfrac{q_{\gamma_{1}}q_{\gamma_{2}}...q_{\gamma_{p_{1}+1}}b(N,\gamma
-\gamma_{1}-\gamma_{2}-...-\gamma_{p_{1}+1})}{\prod_{j=1}^{p_{1}}(\Lambda
_{N}-\mid\gamma+t-\sum_{i=1}^{j}\gamma_{i}\mid^{2l})}.
\]
Here the summations for $S_{k}$ and $C_{p_{1}}$ are taken under the additional
conditions $\gamma_{1}+\gamma_{2}+...+\gamma_{s}\neq0$ for $s=1,2,...,k$ and
$s=1,2,...,p_{1}$ respectively. These conditions and the inclusion $\gamma
_{i}\in\Gamma(\rho^{\alpha})$ for $i=1,2,...,p_{1}$ imply the relation
$\sum_{i=1}^{j}\gamma_{i}\in\Gamma(p\rho^{\alpha})$. Therefore from the second
inequality in (20) it follows that the absolute values of the denominators of
the fractions in $S_{k}$ and $C_{p_{1}}$ are greater than $(\frac{1}{2}%
\rho^{\alpha_{1}})^{k}$ and $(\frac{1}{2}\rho^{\alpha_{1}})^{p_{1}}$
respectively. Hence the first inequality in (4) and $p_{1}\alpha_{1}\geq
p\alpha$ ( see the fourth inequality in (15)) yield
\begin{equation}
C_{p_{1}}=O(\rho^{-p_{1}\alpha_{1}})=O(\rho^{-p\alpha}),\text{ }S_{k}%
(\Lambda_{N},\gamma+t)=O(\rho^{-k\alpha_{1}}),\forall k=1,2,...,p_{1}.
\end{equation}
Since we used only the condition (6) for $\Lambda_{N},$ it follows that
\begin{equation}
S_{k}(a,\gamma+t)=O(\rho^{-k\alpha_{1}})
\end{equation}
for all $a\in\mathbb{R}$ satisfying $\mid a-\mid\gamma+t\mid^{2l}\mid<\frac
{1}{2}\rho^{\alpha_{1}}.$ Thus finding $N$ such that $\Lambda_{N}$ is close to
$\mid\gamma+t\mid^{2l}$ and $b(N,\gamma)$ is not very small, then dividing
both sides of (21) by $b(N,\gamma),$ we get the asymptotic formulas for
$\Lambda_{N}$.

\begin{theorem}
$(a)$ Suppose $\gamma+t\in U^{l}(\rho^{\alpha_{1}},p).$ If (6) and (7) hold,
then $\Lambda_{N}$ satisfies formulas (5) for $k=1,2,...,[\frac{1}{3}(p-c)],$
where
\begin{equation}
F_{s}=O(\rho^{-\alpha_{1}}),\forall s=0,1,...,
\end{equation}
and $F_{0}=0,$ $F_{s}=A_{s}(\mid\gamma+t\mid^{2l}+F_{s-1},\gamma+t)$ for
$s=1,2,....$

$(b)$ For $\gamma+t\in U^{l}(\rho^{\alpha_{1}},p)$ there exists an eigenvalue
$\Lambda_{N}$ of $L_{t}(l,q(x))$ satisfying (5).
\end{theorem}

\begin{proof}
$(a)$ To prove (5) in case $k=1$ we divide both side of (21) by $b(N,\gamma)$
and use (7), (22). Then we obtain
\begin{equation}
\Lambda_{N}-\mid\gamma+t\mid^{2l}=O(\rho^{-\alpha_{1}}).
\end{equation}
This and $\alpha_{1}=3\alpha$ ( see the end of the introduction) imply that
formula (5) for $k=1$ holds and $F_{0}=0.$ Hence (24) for $s=0$ is also
proved. Moreover, from (23), we obtain $S_{k}(\mid\gamma+t\mid^{2l}%
+O(\rho^{-\alpha_{1}}),\gamma+t)=O(\rho^{-\alpha_{1}})$ for $k=1,2,....$
Therefore (24) for arbitrary $s$ follows from the definition of $F_{s}$ by
induction. Now we prove (5) by induction on $k$. Suppose (5) holds for $k=j$,
that is,

$\Lambda_{N}=\mid\gamma+t\mid^{2l}+F_{k-1}(\gamma+t)+O(\rho^{-3k\alpha}).$
Substituting this into $A_{p_{1}}(\Lambda_{N},\gamma+t)$ in (21) and dividing
both sides of (21) by $b(N,\gamma),$ we get
\begin{align*}
\Lambda_{N}  &  =\mid\gamma+t\mid^{2l}+A_{p_{1}}(\mid\gamma+t\mid^{2l}%
+F_{j-1}+O(\rho^{-j\alpha_{1}}),\gamma+t)+O(\rho^{-(p-c)\alpha})=\\
&  \mid\gamma+t\mid^{2l}+\{A_{p_{1}}(\mid\gamma+t\mid^{2l}+F_{j-1}%
+O(\rho^{-j\alpha_{1}}),\gamma+t)-\\
A_{p_{1}}(  &  \mid\gamma+t\mid^{2l}+F_{j-1},\gamma+t)\}+A_{p_{1}}(\mid
\gamma+t\mid^{2l}+F_{j-1},\gamma+t)+O(\rho^{-(p-c)\alpha}).
\end{align*}
To prove $(a)$ for $k=j+1$ we need to show that the expression in curly
brackets is equal to $O(\rho^{-(j+1)\alpha_{1}}).$ It can be checked by using
(4), (20), (24) and the obvious relation
\begin{align*}
&  \frac{1}{\prod_{j=1}^{s}(\mid\gamma+t\mid^{2l}+F_{j-1}+O(\rho^{-j\alpha
_{1}})-\mid\gamma+t-\sum_{i=1}^{s}\gamma_{i}\mid^{2l})}-\\
&  \frac{1}{\prod_{j=1}^{s}(\mid\gamma+t\mid^{2l}+F_{j-1}-\mid\gamma
+t-\sum_{i=1}^{s}\gamma_{i}\mid^{2l})}\\
&  =\frac{1}{\prod_{j=1}^{s}(\mid\gamma+t\mid^{2l}+F_{j-1}-\mid\gamma
+t-\sum_{i=1}^{s}\gamma_{i}\mid^{2l})}(\frac{1}{1-O(\rho^{-(j+1)\alpha_{1}}%
)}-1)
\end{align*}
$=O(\rho^{-(j+1)\alpha_{1}})$ for $s=1,2,...,p_{1}.$

$(b)$ Let $A$ be the set of indices $N$ satisfying (6). Using (16) and Bessel
inequality, we obtain%
\[
\sum_{N\notin A}\mid b(N,\gamma)\mid^{2}=\sum_{N\notin A}\mid\dfrac{(\Psi
_{N}(x),q(x)e^{i(\gamma+t,x)})}{\Lambda_{N}-\mid\gamma+t\mid^{2l}}\mid
^{2}=O(\rho^{-2\alpha_{1}})
\]
Hence, by the Parseval equality, we have $\sum_{N\in A}\mid b(N,\gamma
)\mid^{2}=1-O(\rho^{-2\alpha_{1}}).$ This and the inequality $\mid A\mid
<c_{5}\rho^{d-1}=c_{5}\rho^{(d-1)m\alpha}$ ( see the end of the introduction)
imply that there exists a number $N$ satisfying $\mid b(N,\gamma)\mid>\frac
{1}{2}(c_{5})^{-1}\rho^{-\frac{(d-1)m}{2}\alpha}$, that is, (7) holds for
$c=\frac{(d-1)m}{2}$ . Thus $\Lambda_{N}$ satisfies (5) due to $(a)$
\end{proof}

Theorem 1 shows that in the non-resonance case the eigenvalue of the perturbed
operator $L_{t}(l,q(x))$ is close to the eigenvalue of the unperturbed
operator $L_{t}(l,0).$ However, in Theorem 2 we prove that if $\gamma+t\in
\cap_{i=1}^{k}V_{\gamma_{i}}^{1}(\rho^{\alpha_{k}})\backslash E_{k+1}^{1}$ for
$k\geq1,$ where $\gamma_{1},\gamma_{2},...,\gamma_{k}$ are linearly
independent vectors of $\Gamma(p\rho^{\alpha}),$ then the corresponding
eigenvalue of $L_{t}(l,q(x))$ is close to the eigenvalue of the matrix
constructed as follows. Introduce the sets:

$B_{k}\equiv B_{k}(\gamma_{1},\gamma_{2},...,\gamma_{k})=\{b:b=\sum_{i=1}%
^{k}n_{i}\gamma_{i},n_{i}\in Z,\mid b\mid<\frac{1}{2}\rho^{\frac{1}{2}%
\alpha_{k+1}}\},$%
\begin{equation}
B_{k}(\gamma+t)=\gamma+t+B_{k}=\{\gamma+t+b:b\in B_{k}\},
\end{equation}
$B_{k}(\gamma+t,p_{1})=\{\gamma+t+b+a:b\in B_{k},\mid a\mid<p_{1}\rho^{\alpha
},a\in\Gamma\}.$

Denote by $h_{i}+t$ for $i=1,2,...,b_{k}$ the vectors of $B_{k}(\gamma
+t,p_{1}),$ where

$b_{k}\equiv b_{k}(\gamma_{1},\gamma_{2},...,\gamma_{k})$ is the number of the
vectors of $B_{k}(\gamma+t,p_{1})$. Define the matrix $C(\gamma+t,\gamma
_{1},\gamma_{2},...,\gamma_{k})\equiv(c_{i,j})$ by the formulas
\begin{equation}
c_{i,i}=\mid h_{i}+t\mid^{2l},\text{ }c_{i,j}=q_{h_{i}-h_{j}},\text{ }\forall
i\neq j,
\end{equation}
where $i,j=1,2,...,b_{k}.$

Using the mean value theorem it is not hard to see that if

$x\in\mathbb{R}^{d},$ $\mid x\mid\sim\rho,$ $\gamma_{1}\in\Gamma,$ $\mid
x+\gamma_{1}\mid\sim\rho,$ then
\begin{equation}
\mid x\mid^{2l}-\mid x+\gamma_{1}\mid^{2l}=a^{2(l-1)}(\mid x\mid^{2}-\mid
x+\gamma_{1}\mid^{2})
\end{equation}
where $a\sim\rho.$ Therefore for $l\geq1$ and $k=1,2,...$ ,\ we have%
\begin{equation}
(\cap_{i=1}^{k}V_{\gamma_{i}}^{l}(\rho^{\alpha_{k}}))\subset\cap_{i=1}%
^{k}V_{\gamma_{i}}^{1}(\rho^{\alpha_{k}})),
\end{equation}%
\begin{equation}
U^{1}(\rho^{\alpha_{1}},p)\subset U^{l}(\rho^{\alpha_{1}},p)
\end{equation}
Taking into account this, we consider the resonance eigenvalue \ $\mid
\gamma+t\mid^{2l}$ for $\gamma+t\in(\cap_{i=1}^{k}V_{\gamma_{i}}^{1}%
(\rho^{\alpha_{k}}))$ by using the following lemma.

\begin{lemma}
If $\gamma+t\in\cap_{i=1}^{k}V_{\gamma_{i}}^{1}(\rho^{\alpha_{k}})\backslash
E_{k+1}^{1},$ $h+t\in B_{k}(\gamma+t,p_{1}),$

$(h-\gamma^{^{\prime}}+t)\notin B_{k}(\gamma+t,p_{1}),$ then
\begin{equation}
\mid\mid\gamma+t\mid^{2l}-\mid h-\gamma^{^{\prime}}-\gamma_{1}^{^{\prime}%
}-\gamma_{2}^{^{\prime}}-...-\gamma_{s}^{^{\prime}}+t\mid^{2l}\mid>\frac{1}%
{5}\rho^{\alpha_{k+1}},
\end{equation}
where $\gamma^{^{\prime}}\in\Gamma(\rho^{\alpha}),$ $\gamma_{j}^{^{\prime}}%
\in\Gamma(\rho^{\alpha}),$ $j=1,2,...,s$ and $s=0,1,...,p_{1}-1.$
\end{lemma}

\begin{proof}
The inequality $p>2p_{1}$ ( see the end of the introduction) and the
conditions of Lemma 1 imply that

$h-\gamma^{^{\prime}}-\gamma_{1}^{^{\prime}}-\gamma_{2}^{^{\prime}}%
-...-\gamma_{s}^{^{\prime}}+t\in B_{k}(\gamma+t,p)\backslash B_{k}(\gamma+t)$
for all $s=0,1,...,p_{1}-1.$ It follows from the definitions of $B_{k}%
(\gamma+t,p),$ $B_{k}$ that ( see (26))

$h-\gamma^{^{\prime}}-\gamma_{1}^{^{\prime}}-\gamma_{2}^{^{\prime}}%
-...-\gamma_{s}^{^{\prime}}+t=\gamma+t+b+a,$ where
\begin{equation}
\mid b\mid<\frac{1}{2}\rho^{\frac{1}{2}\alpha_{k+1}},\mid a\mid<p\rho^{\alpha
},\text{ }\gamma+t+b+a\notin\gamma+t+B_{k}.
\end{equation}
Then (31) has the form
\begin{equation}
\mid\mid\gamma+t+a+b\mid^{2l}-\mid\gamma+t\mid^{2l}\mid>\frac{1}{5}%
\rho^{\alpha_{k+1}}.
\end{equation}
It follows from (28) that, to verify (33) it is enough to prove it for $l=1.$
To prove (33) for $l=1$ we consider two cases:

Case 1. $a\in P$, where $P=Span\{\gamma_{1,}\gamma_{2},...,\gamma_{k}\}.$
Since $b\in B_{k}\subset P,$ we have $a+b\in P.$ This with the third relation
in (32) imply that $a+b\in P\backslash B_{k}$ ,i.e., $\mid a+b\mid\geq\frac
{1}{2}$ $\rho^{\frac{1}{2}\alpha_{k+1}}$. Consider the orthogonal
decomposition $\gamma+t=y+v$ of $\gamma+t,$ where $v\in P$ and $y\bot P.$
First we prove that the projection $v$ of any vector $x\in\cap_{i=1}%
^{k}V_{\gamma_{i}}^{1}(\rho^{\alpha_{k}})$ on $P$ satisfies
\begin{equation}
\mid v\mid=O(\rho^{(k-1)\alpha+\alpha_{k}}).
\end{equation}
For this we turn the coordinate axis so that $Span\{\gamma_{1,}\gamma
_{2},...,\gamma_{k}\}$ coincides with the span of the vectors $e_{1}%
=(1,0,0,...,0)$, $e_{2}=(0,1,0,...,0),...,$ $e_{k}$. Then $\gamma_{s}%
=\sum_{i=1}^{k}\gamma_{s,i}e_{i}$ for $s=1,2,...,k$ . Therefore the relation
$x\in\cap_{i=1}^{k}V_{\gamma_{i}}^{1}(\rho^{\alpha_{k}})$ implies that
\[
\sum_{i=1}^{k}\gamma_{s,i}x_{i}=O(\rho^{\alpha_{k}}),s=1,2,...,k;\text{ }%
x_{n}=\frac{\det(b_{j,i}^{n})}{\det(\gamma_{j,i})}\text{, }n=1,2,...,k,
\]
where $x=(x_{1},x_{2},...,x_{d}),\gamma_{j}=(\gamma_{j,1},\gamma
_{j,2},...,\gamma_{j,k},0,0,...,0),$ $b_{j,i}^{n}=\gamma_{j,i}$ for $n\neq j$
and $b_{j,i}^{n}=O(\rho^{\alpha_{k}})$ for $n=j.$ Taking into account that the
determinant $\det(\gamma_{j,i})$ is the volume of the parallelepiped
$\{\sum_{i=1}^{k}b_{i}\gamma_{i}:b_{i}\in\lbrack0,1],i=1,2,...,k\}$ and using
$\mid\gamma_{j,i}\mid<p\rho^{\alpha}$ ( since $\gamma_{j}\in\Gamma
(p\rho^{\alpha})$ ), we get the estimations
\begin{equation}
x_{n}=O(\rho^{\alpha_{k}+(k-1)\alpha})\text{ ,}\forall n=1,2,...,k;\text{
}\forall x\in\cap_{i=1}^{k}V_{\gamma_{i}}^{1}(\rho^{\alpha_{k}}).
\end{equation}
Hence (34) holds. Therefore, using the inequalities $\mid a+b\mid\geq\frac
{1}{2}$ $\rho^{\frac{1}{2}\alpha_{k+1}}$ ( see above), $\alpha_{k+1}%
>2(\alpha_{k}+(k-1)\alpha)$ ( see the \ seventh inequality in (15)), and the
obvious equalities $(y,v)=(y,a)=(y,b)=0,$%
\begin{equation}
\mid\gamma+t+a+b\mid^{2}-\mid\gamma+t\mid^{2}=\mid a+b+v\mid^{2}-\mid
v\mid^{2},
\end{equation}
we obtain the estimation (33).

Case 2. $a\notin P.$ First we show that
\begin{equation}
\mid\mid\gamma+t+a\mid^{2}-\mid\gamma+t\mid^{2}\mid\geq\rho^{\alpha_{k+1}}.
\end{equation}
Suppose, to the contrary, that it does not hold. Then $\gamma+t\in V_{a}%
^{1}(\rho^{\alpha_{k+1}}).$ On the other hand $\gamma+t\in\cap_{i=1}%
^{k}V_{\gamma_{i}}^{1}(\rho^{\alpha_{k+1}})$ ( see the conditions of Lemma 1).
Therefore we have $\gamma+t\in E_{k+1}^{1}$ which contradicts the conditions
of the lemma. \ Hence (37) is proved. Now, to prove (33) we write the
difference $\mid\gamma+t+a+b\mid^{2}-\mid\gamma+t\mid^{2}$ as the sum of
$d_{1}\equiv\mid\gamma+t+a+b\mid^{2}-\mid\gamma+t+b\mid^{2}$ and $d_{2}%
\equiv\mid\gamma+t+b\mid^{2}-\mid\gamma+t\mid^{2}.$ Since $d_{1}=\mid
\gamma+t+a\mid^{2}-\mid\gamma+t\mid^{2}+2(a,b),$ it follows from the
inequalities (37), (32) that $\mid d_{1}\mid>\frac{2}{3}$ $\rho^{\alpha_{k+1}%
}$. On the other hand, taking $a=0$ in (36), we have $d_{2}=\mid b+v\mid
^{2}-\mid v\mid^{2}.$ Therefore (34), the first inequality in (32) and the
\ seventh inequality in (15) imply that $\mid d_{2}\mid<\frac{1}{3}$
$\rho^{\alpha_{k+1}},$ $\mid d_{1}\mid-\mid d_{2}\mid>\frac{1}{3}\rho
^{\alpha_{k+1}},$ that is, (33) holds
\end{proof}

\begin{theorem}
$(a)$ Suppose $\gamma+t\in(\cap_{i=1}^{k}V_{\gamma_{i}}^{1}(\rho^{\alpha_{k}%
}))\backslash E_{k+1}^{1},$ where $k=1,2,...,d-1.$ If (6) and (7) hold, then
there is an index $j$ such that
\begin{equation}
\Lambda_{N}(t)=\lambda_{j}(\gamma+t)+O(\rho^{-(p-c-\frac{1}{4}d3^{d})\alpha}),
\end{equation}
where $\lambda_{1}(\gamma+t)\leq\lambda_{2}(\gamma+t)\leq...\leq\lambda
_{b_{k}}(\gamma+t)$ are the eigenvalues of the matrix $C(\gamma+t,\gamma
_{1},\gamma_{2},...,\gamma_{k})$ defined in (27).

$(b)$ Every eigenvalue $\Lambda_{N}(t)$ of the operator $L_{t}(l,q(x))$
satisfies either (5) or (38) for $c=\frac{m(d-1)}{2}.$
\end{theorem}

\begin{proof}
$(a)$Writing the equation (17) for all $h_{i}+t\in B_{k}(\gamma+t,p_{1}),$ we
obtain%
\begin{equation}
(\Lambda_{N}-\mid h_{i}+t\mid^{2l})b(N,h_{i})=\sum_{\gamma^{^{\prime}}%
\in\Gamma(\rho^{\alpha})}q_{\gamma^{^{\prime}}}b(N,h_{i}-\gamma^{^{\prime}%
})+O(\rho^{-p\alpha})
\end{equation}
for $i=1,2,...,b_{k}$ ( see (26) for definition of $B_{k}(\gamma+t,p_{1})$).
It follows from (6) and Lemma 1 that if $(h_{i}-\gamma^{^{\prime}}+t)\notin
B_{k}(\gamma+t,p_{1}),$ then
\[
\mid\Lambda_{N}-\mid h_{i}-\gamma^{^{\prime}}-\gamma_{1}-\gamma_{2}%
-...-\gamma_{s}+t\mid^{2l}\mid>\frac{1}{6}\rho^{\alpha_{k+1}},
\]
where $\gamma^{^{\prime}}\in\Gamma(\rho^{\alpha}),\gamma_{j}\in\Gamma
(\rho^{\alpha}),$ $j=1,2,...,s$ and $s=0,1,...,p_{1}-1.$ Therefore, applying
the formula (18) $p_{1}$ times, using (4) and $p_{1}\alpha_{k+1}>p_{1}%
\alpha_{1}\geq p\alpha$ ( see the\ fourth inequality in (15)), we see that if
$(h_{i}-\gamma^{^{\prime}}+t)\notin B_{k}(\gamma+t,p_{1}),$ then
\begin{equation}
b(N,h_{i}-\gamma^{^{\prime}})=\nonumber
\end{equation}%
\begin{equation}
\sum_{\gamma_{1},...,\gamma_{p_{1}-1}\in\Gamma(\rho^{\alpha})}\dfrac
{q_{\gamma_{1}}q_{\gamma_{2}}...q_{\gamma_{p_{1}}}b(N,h_{i}-\gamma^{^{\prime}%
}-\sum_{i=1}^{p_{1}}\gamma_{i})}{\prod_{j=0}^{p_{1}-1}(\Lambda_{N}-\mid
h_{i}-\gamma^{^{\prime}}+t-\sum_{i=1}^{j}\gamma_{i}\mid^{2l})}+
\end{equation}%
\[
+O(\rho^{-p\alpha})=O(\rho^{p_{1}\alpha_{k+1}})+O(\rho^{-p\alpha}%
)=O(\rho^{-p\alpha}).
\]
Hence (39) has the form%
\[
(\Lambda_{N}-\mid h_{i}+t\mid^{2l})b(N,h_{i})=\sum_{\gamma^{^{\prime}}%
}q_{\gamma^{^{\prime}}}b(N,h_{i}-\gamma^{^{\prime}})+O(\rho^{-p\alpha
}),i=1,2,...,b_{k},
\]
where the summation is taken under the conditions $\gamma^{^{\prime}}\in
\Gamma(\rho^{\alpha})$ and

$h_{i}-\gamma^{^{\prime}}+t\in B_{k}(\gamma+t,p_{1})$. It can be written in
matrix form
\[
(C-\Lambda_{N}I)(b(N,h_{1}),b(N,h_{2}),...b(N,h_{b_{k}}))=O(\rho^{-p\alpha}),
\]
where the right-hand side of this system is a vector having the norm

$\parallel O(\rho^{-p\alpha})\parallel=O(\sqrt{b_{k}}\rho^{-p\alpha})$. Now,
taking into account that

$\gamma+t\in\{h_{i}+t:i=1,2,...,b_{k}\}$ and (7) holds, we have
\begin{align}
c_{4}\rho^{-c\alpha}  &  <(\sum_{i=1}^{b_{k}}\mid b(N,h_{i})\mid^{2}%
)^{\frac{1}{2}}\leq\parallel(C-\Lambda_{N}I)^{-1}\parallel\sqrt{b_{k}}%
c_{6}\rho^{-p\alpha},\\
\max_{i=1,2,...,b_{k}}  &  \mid\Lambda_{N}-\lambda_{i}\mid^{-1}=\parallel
(C-\Lambda_{N}I)^{-1}\parallel>c_{4}c_{6}^{-1}b_{k}^{-\frac{1}{2}}%
\rho^{-c\alpha+p\alpha}.
\end{align}
Since $b_{k}$ is the number of the vectors of $B_{k}(\gamma+t,p_{1}),$ it
follows from the definition of $B_{k}(\gamma+t,p_{1})$ ( see (26)) and the
obvious relations $\mid B_{k}\mid=O(\rho^{\frac{k}{2}\alpha_{k+1}}),$

$\mid\Gamma(p_{1}\rho^{\alpha})\mid=O(\rho^{d\alpha})$ and $d\alpha<\frac
{1}{2}\alpha_{d}$ ( see the end of introduction), we get
\begin{equation}
b_{k}=O(\rho^{d\alpha+\frac{k}{2}\alpha_{k+1}})=O(\rho^{\frac{d}{2}\alpha_{d}%
})=O(\rho^{\frac{d}{2}3^{d}\alpha}),\forall k=1,2,...,d-1
\end{equation}
Thus formula (38) follows from (42) and (43).

$(b)$ Let $\Lambda_{N\text{ }}(t)$ be any eigenvalue of the operator
$L_{t}(l,q(x))$ such that

$\sqrt[2l]{\Lambda_{N}(t)}\in(\frac{3}{4}\rho,\frac{5}{4}\rho).$ Denote by $D$
the set of all vectors $\gamma\in\Gamma$ satisfying (6). From (16), arguing as
in the proof of Theorem 1($b$), we obtain

$\sum_{\gamma\in D}\mid b(N,\gamma)\mid^{2}=1-O(\rho^{-2\alpha_{1}}).$ Since
$\mid D\mid=O(\rho^{d-1})$ ( see the end of the introduction), there exists
$\gamma\in D$ such that

$\mid b(N,\gamma)\mid>c_{7}\rho^{-\frac{(d-1)}{2}}=c_{7}\rho^{-\frac
{(d-1)m}{2}\alpha}$, that is, condition (7) for $c=\frac{(d-1)m}{2}$ holds.
Now the proof of $(b)$ follows from Theorem 1 $(a)$ and Theorem 2$(a),$ since
either $\gamma+t$ $\in U^{1}(\rho^{\alpha_{1}},p)$ or $\gamma+t\in$ $E_{k}%
^{1}\backslash E_{k+1}^{1}$ for $k=1,2,...,d-1$ ( see (46))
\end{proof}

\begin{remark}
Here we note that the non-resonance domain

$U^{l}(c_{8}\rho^{\alpha_{1}},p)=(R(\frac{3}{2}\rho)\backslash R(\frac{1}%
{2}\rho))\backslash\bigcup_{\gamma_{1}\in\Gamma(p\rho^{\alpha})}V_{\gamma_{1}%
}^{l}(c_{8}\rho^{\alpha_{1}}),$ where

$V_{\gamma_{1}}^{l}(c_{8}\rho^{\alpha_{1}})\equiv\{x:\mid\mid x\mid^{2l}-\mid
x+\gamma_{1}\mid^{2l}\mid<c_{8}\rho^{\alpha_{1}}\}\cap(R(\frac{3}{2}%
\rho)\backslash R(\frac{1}{2}\rho)),$ has an asymptotically full measure on
$\mathbb{R}^{d}$ in the sense that $\frac{\mu(U^{l}\cap B(\rho))}{\mu
(B(\rho))}$ tends to $1$ as $\rho$ tends to infinity, where $B(\rho
)=\{x\in\mathbb{R}^{d}:\mid x\mid=\rho\}.$ By (30) it is enough to prove this
for $l=1.$ Clearly, $B(\rho)\cap V_{b}^{1}(c_{8}\rho^{\alpha_{1}})$ is the
part of sphere $B(\rho),$ which is contained between two parallel hyperplanes

$\{x:\mid x\mid^{2}-\mid x+b\mid^{2}=-c_{8}\rho^{\alpha_{1}}\}$ and $\{x:\mid
x\mid^{2}-\mid x+b\mid^{2}=c_{8}\rho^{\alpha_{1}}\}.$ The distance of these
hyperplanes from origin is $O(\frac{\rho^{\alpha_{1}}}{\mid b\mid}).$
Therefore, the relations $\mid\Gamma(p\rho^{\alpha})\mid=O(\rho^{d\alpha}),$
and $\alpha_{1}+d\alpha<1-\alpha$ ( see the first inequality in (15)) imply
\begin{align}
\mu(B(\rho)\cap V_{b}^{1}(c_{8}\rho^{\alpha_{1}}))  &  =O(\frac{\rho
^{\alpha_{1}+d-2}}{\mid b\mid}),\text{ }\mu(E_{1}^{1}\cap B(\rho
))=O(\rho^{d-1-\alpha}),\\
\mu(U^{1}(c_{8}\rho^{\alpha_{1}},p)\cap B(\rho))  &  =(1+O(\rho^{-\alpha}%
))\mu(B(\rho)).
\end{align}
If $x\in\cap_{i=1}^{d}V_{\gamma_{i}}^{1}(\rho^{\alpha_{d}}),$ then (35) holds
for $k=d$ and $n=1,2,...,d.$ Hence we have $\mid x\mid=O(\rho^{\alpha
_{d}+(d-1)\alpha}).$ It is impossible, since $\alpha_{d}+(d-1)\alpha<1$ ( see
the \ sixth inequality in (15)) and $x\in B(\rho).$ It means that $(\cap
_{i=1}^{d}V_{\gamma_{i}}^{1}(\rho^{\alpha_{k}}))\cap B(\rho)=\emptyset$ for
$\rho\gg1$. Thus for $\rho\gg1$ we have
\begin{equation}
R(\frac{3}{2}\rho)\backslash R(\frac{1}{2}\rho)=(U^{1}(\rho^{\alpha_{1}%
},p)\cup(\cup_{s=1}^{d-1}(E_{s}^{1}\backslash E_{s+1}^{1}))).
\end{equation}

Note that everywhere in this paper we use the big parameter $\rho.$ All
considered eigenvalues $\left\vert \gamma+t\right\vert ^{2l}$ of $L_{t}(l,0)$
satisfy the relations $\frac{1}{2}\rho<\left\vert \gamma+t\right\vert
<\frac{3}{2}\rho.$ Therefore in the asymptotic formulas instead of $O(\rho
^{a})$ one can take \ $O(\left\vert \gamma+t\right\vert ^{a}).$ For
simplicity, we often use $O(\rho^{a}).$ It is clear that the asymptotic
formulas hold true if we replace $U^{l}(\rho^{\alpha_{1}},p)$ by $U^{l}%
(c_{8}\rho^{\alpha_{1}},p),$ where instead of $c_{8}$ one can write $\frac
{1}{2}$ or $\frac{3}{2}.$ Since $V_{b}^{l}(\frac{1}{2}\rho^{\alpha_{1}%
})\subset(R(\frac{3}{2}\rho)\backslash R(\frac{1}{2}\rho))\cap W_{b,\alpha
_{1}}^{l}(1)\subset$ $V_{b}^{l}(\frac{3}{2}\rho^{\alpha_{1}}),$\ in all
considerations the resonance domain $V_{b}^{l}(\rho^{\alpha_{1}})$ can be
replaced by $W_{b,\alpha_{1}}^{l}(1)\cap(R(\frac{3}{2}\rho)\backslash
R(\frac{1}{2}\rho)).$
\end{remark}

\begin{remark}
Here we note some properties of the known part

$\mid\gamma+t\mid^{2l}+F_{k}(\gamma+t)$ (see Theorem 1) of the non-resonance
eigenvalues of $L_{t}(l,q(x)).$ Denoting $\gamma+t$ by $x$ , where
$\gamma+t\in U^{1}(\rho^{\alpha_{1}},p),$ we prove
\begin{equation}
\frac{\partial F_{k}(x)}{\partial x_{i}}=O(\rho^{2-2l-2\alpha_{1}+\alpha
}),\forall i=1,2,...,d;\forall k=1,2,...
\end{equation}
by induction on $k.$ Using (28) one can easily verify that if $\mid x\mid
\sim\rho,$ and

$x\in U^{1}(\rho^{\alpha_{1}},p),$ that is, if $x\notin V_{\gamma_{1}}%
^{1}(\rho^{\alpha_{1}})$ for $\gamma_{1}\in\Gamma(p\rho^{\alpha}),$ then
\begin{align*}
&  \mid x\mid^{2l}-\mid x-\gamma_{1}\mid^{2l}\sim\rho^{2l-2}(\mid x\mid
^{2}-\mid x-\gamma_{1}\mid^{2}),\\
&  \mid x\mid^{2l-2}-\mid x-\gamma_{1}\mid^{2l-2}\sim\rho^{2l-4}(\mid
x\mid^{2}-\mid x-\gamma_{1}\mid^{2}),
\end{align*}

\ \ \ \ \ \ \ $\ \ \ \ \ \ \ \ \ \ \ \ \ \ \ \ \ \mid\mid x\mid^{2}-\mid
x-\gamma_{1}\mid^{2}\mid>\rho^{\alpha_{1}},$

$\ \ \ \ \ \ \ \ \ \ \ \ \ \ \ \ \ \ \ \ \ \ \ \ \ \frac{\partial}{\partial
x_{i}}(\dfrac{1}{\mid x\mid^{2l}-\mid x-\gamma_{1}\mid^{2l}})=$%
\begin{equation}
-\dfrac{2lx_{i}(\mid x\mid^{2l-2}-\mid x-\gamma_{1}\mid^{2l-2})}{(\mid
x\mid^{2l}-\mid x-\gamma_{1}\mid^{2l})^{2}})+
\end{equation}%
\[
\dfrac{2\gamma_{1}(i)\mid x-\gamma_{1}\mid^{2l-2}}{(\mid x\mid^{2l}-\mid
x-\gamma_{1}\mid^{2l})^{2}}=O(\rho^{2-2l-2\alpha_{1}+\alpha}),
\]
where $(\gamma_{1}(1),\gamma_{1}(2),...,\gamma_{1}(d))=\gamma_{1}\in
\Gamma(p\rho^{\alpha})$ and hence $\gamma_{1}(i)=O(\rho^{\alpha}).$ Now (47)
for $k=1$ follows from (4) and (48). Suppose that (47) holds for $k=s.$ Using
this and (24), replacing $\mid x\mid^{2l}$ by $\mid x\mid^{2l}+F_{s}(x)$ in
(48), and evaluating as above, we obtain%
\[
\frac{\partial}{\partial x_{i}}(\dfrac{1}{\mid x\mid^{2l}+F_{s}(x)-\mid
x-\gamma_{1}\mid^{2l}})=
\]%
\[
O(\rho^{2l-2-2\alpha_{1}+\alpha})-\dfrac{\frac{\partial F_{s}(x)}{\partial
x_{i}}}{(\mid x\mid^{2l}+F_{s}-\mid x-\gamma_{1}\mid^{2l})^{2}}=
\]%
\[
O(\rho^{2-2l-2\alpha_{1}+\alpha})+O(\rho^{2-2l-4\alpha_{1}+\alpha}%
)=O(\rho^{2-2l-2\alpha_{1}+\alpha})
\]
This formula with the definition of $F_{k}$ implies (47) for $k=s+1.$
\end{remark}

\section{Asymptotic Formulas for Bloch Functions}

\bigskip In this section using the asymptotic formulas for the eigenvalues and
the simplicity conditions (12), (13), we prove the asymptotic formulas for the
Bloch functions with a quasimomentum of the simple set $B$.

\begin{theorem}
If $\gamma+t\in B,$ then there exists a unique eigenvalue $\Lambda_{N}(t)$
satisfying (5) for $k=1,2,...,[\frac{p}{3}],$ where $p$ is defined in (3).
This is a simple eigenvalue and the corresponding eigenfunction $\Psi
_{N,t}(x)$ of $L(l,q(x))$ satisfies (10) if

$q(x)\in W_{2}^{s_{0}}(F),$ where $s_{0}=\frac{3d-1}{2}(3^{d}+d+2)+\frac{1}%
{4}d3^{d}+d+6.$
\end{theorem}

\begin{proof}
By Theorem 1(b) if $\gamma+t\in B\subset U^{1}(\rho^{\alpha_{1}},p),$ then
there exists an eigenvalue $\Lambda_{N}(t)$ satisfying (5) for
$k=1,2,...,[\frac{1}{3}(p-\frac{1}{2}m(d-1))].$ Since

$k_{1}=[\frac{d}{3\alpha}]+2\leq\frac{1}{3}(p-\frac{1}{2}m(d-1))$ (see the
third inequality in (15)) formula (5) holds for $k=k_{1}.$ Therefore using
(5), the relation $3k_{1}\alpha>d+2\alpha$ ( see the fifth inequality in
(15)), and notations $F(\gamma+t)=\mid\gamma+t\mid^{2l}+F_{k_{1}-1}(\gamma
+t)$, $\varepsilon_{1}=\rho^{-d-2\alpha}$ ( see Step 1 in introduction), we
obtain
\begin{equation}
\Lambda_{N}(t)=F(\gamma+t)+o(\varepsilon_{1}).
\end{equation}
Let $\Psi_{N}$ be any normalized eigenfunction corresponding to $\Lambda_{N}$.
Since the normalized eigenfunction is defined up to constant of modulus $1,$
without loss of generality it can assumed that $\arg b(N,\gamma)=0,$ where
$b(N,\gamma)=(\Psi_{N},e^{i(\gamma+t,x)}).$ Therefore to prove (10) it
suffices to show that (14) holds. To prove (14) first \ we estimate
\ $\sum_{\gamma^{^{\prime}}\notin K}\mid b(N,\gamma^{^{\prime}})\mid^{2}$and
then $\sum_{\gamma^{^{\prime}}\in K\backslash\{\gamma\}}\mid b(N,\gamma
^{^{\prime}})\mid^{2},$ where $K$ is defined in (12), (13). Using (102), the
definition of $K$, and (16), we get
\begin{align}
&  \mid\Lambda_{N}-\mid\gamma^{^{\prime}}+t\mid^{2l}\mid>\frac{1}{4}%
\rho^{\alpha_{1}},\text{ }\forall\gamma^{^{\prime}}\notin K,\\
\sum_{\gamma^{^{\prime}}\notin K}  &  \mid b(N,\gamma^{^{\prime}})\mid
^{2}=\parallel q(x)\Psi_{N}\parallel^{2}O(\rho^{-2\alpha_{1}})=O(\rho
^{-2\alpha_{1}}).\nonumber
\end{align}
If $\gamma^{^{\prime}}\in K$ , then by (49) and by definition of $K,$ it
follows that
\begin{equation}
\mid\Lambda_{N}-\mid\gamma^{^{\prime}}+t\mid^{2l}\mid<\frac{1}{2}\rho
^{\alpha_{1}}%
\end{equation}
Now we prove that the simplicity conditions (12), (13) imply
\begin{equation}
\mid b(N,\gamma^{^{\prime}})\mid\leq c_{4}\rho^{-c\alpha},\text{ }%
\forall\gamma^{^{\prime}}\in K\backslash\{\gamma\},
\end{equation}
where $c=p-dm-\frac{1}{4}d3^{d}-3.$ The conditions $\gamma^{^{\prime}}\in K,$
$\gamma+t\in B$ and (24) imply the inclusion $\gamma^{^{\prime}}+t\in
R(\frac{3}{2}\rho)\backslash R(\frac{1}{2}\rho).$ If for $\gamma^{^{\prime}%
}+t\in U^{1}(\rho^{\alpha_{1}},p)$ and $\gamma^{^{\prime}}\in K\backslash
\{\gamma\}$ the inequality in (52) is not true, then by (51) and Theorem 1(a),
we have
\begin{equation}
\Lambda_{N}=\mid\gamma^{^{\prime}}+t\mid^{2l}+F_{k-1}(\gamma^{^{\prime}%
}+t)+O(\rho^{-3k\alpha})
\end{equation}
for $k=1,2,...,[\frac{1}{3}(p-c)]=[\frac{1}{3}(dm+\frac{1}{4}d3^{d}+3)].$
Since $\alpha=\frac{1}{m}$ and

$k_{1}\equiv\lbrack\frac{d}{3\alpha}]+2<\frac{1}{3}(dm+\frac{1}{4}d3^{d}+3)$,
\ the formula (53) holds for $k=k_{1}.$ Therefore arguing as in the prove of
(49), we get $\Lambda_{N}-F(\gamma^{^{\prime}}+t)=o(\varepsilon_{1})$. \ This
with (49) contradicts (12). Similarly, if the inequality in (52) does not hold
for $\gamma^{^{\prime}}+t\in(E_{k}^{1}\backslash E_{k+1}^{1})$ and
$\gamma^{^{\prime}}\in K,$ then by Theorem 2(a)
\begin{equation}
\Lambda_{N}=\lambda_{j}(\gamma^{^{\prime}}+t)+O(\rho^{-(p-c-\frac{1}{4}%
d3^{d})\alpha}),
\end{equation}
where $(p-c-\frac{1}{4}d3^{d})\alpha=(dm+3)\alpha>d+2\alpha$ . Hence we have

$\Lambda_{N}-\lambda_{j}(\gamma^{^{\prime}}+t)=o(\varepsilon_{1}).$ This with
(49) contradicts (13). So the inequality in (52) holds. Therefore, using $\mid
K\mid=O(\rho^{d-1}),$ $m\alpha=1,$ we get
\begin{equation}
\sum_{\gamma^{^{\prime}}\in K\backslash\{\gamma\}}\mid b(N,\gamma^{^{\prime}%
})\mid^{2}=O(\rho^{-(2c-q(d-1))\alpha})=O(\rho^{-(2p-(3d-1)q-\frac{1}{2}%
d3^{d}-6)\alpha}).
\end{equation}
If $s=s_{0},$ that is, $p=s_{0}-d,$ then $2p-(3d-1)m-\frac{1}{2}d3^{d}-6=6.$
Since $\alpha_{1}=3\alpha,$ the equality (55) and the equality in (50) imply
(14). Thus we proved that the equality (10) holds for any normalized
eigenfunction $\Psi_{N}$ corresponding to any eigenvalue $\Lambda_{N}$
\ satisfying (5). If there exist two different eigenvalues or multiple
eigenvalue satisfying (5), then there exist two orthogonal normalized
eigenfunction satisfying (10), which is impossible. Therefore $\Lambda_{N}%
$\ is a simple eigenvalue. It follows from Theorem 1(a) that $\Lambda_{N}$
satisfies (5) for $k=1,2,...,[\frac{p}{3}],$ since the inequality (7) holds
for $c=0$ ( see (10)).
\end{proof}

\begin{remark}
Since for $\gamma+t\in B$ \ there exists a unique eigenvalue satisfying (5),
(49) we denote this eigenvalue by $\Lambda(\gamma+t).$ Since this eigenvalue
is simple, we denote the \ corresponding eigenfunction by $\Psi_{\gamma
+t}(x).$ By Theorem 3 this eigenfunction satisfies (10). Clearly, for
$\gamma+t\in B$ \ there exists a unique index $N\equiv N(\gamma+t)$ such that
$\Lambda(\gamma+t)=\Lambda_{N(\gamma+t)}$) and $\Psi_{\gamma+t}(x)=\Psi
_{N(\gamma+t)}(x)).$
\end{remark}

Now we prove the asymptotic formulas of arbitrary order for $\Psi_{\gamma
+t}(x).$

\begin{theorem}
If $\gamma+t\in B,$ then the eigenfunction $\Psi_{\gamma+t}(x)\equiv
\Psi_{N(\gamma+t)}(x)$ corresponding to the eigenvalue $\Lambda_{N}%
\equiv\Lambda(\gamma+t)$ satisfies formulas (11), for

$k=1,2,...,n$, where $n=[\frac{1}{6}(2p-(3d-1)m-\frac{1}{2}d3^{d}-6)],$

$\Phi_{0}(x)=0,$ $\Phi_{1}(x)=%
%TCIMACRO{\dsum _{\gamma_{1}\in\Gamma(\rho^{\alpha})}}%
%BeginExpansion
{\displaystyle\sum_{\gamma_{1}\in\Gamma(\rho^{\alpha})}}
%EndExpansion
\dfrac{q_{\gamma_{1}}e^{i(\gamma+t+\gamma_{1},x)}}{(\mid\gamma+t\mid^{2l}%
-\mid\gamma+\gamma_{1}+t\mid^{2l})},$

and $\Phi_{k-1}(x)$ for $k>2$ is a linear combination of $e^{i(\gamma
+t+\gamma^{^{\prime}},x)}$ for

$\gamma^{^{\prime}}\in\Gamma((k-1)\rho^{\alpha})\cup\{0\}$ with coefficients
(61), (62).
\end{theorem}

\begin{proof}
By Theorem 3, formula (11) for $k=1$ is proved. To prove formula (11) for
arbitrary $k\leq n$ we prove the following equivalent relations
\begin{equation}
\sum_{\gamma^{^{\prime}}\in\Gamma^{c}(k-1)}\mid b(N,\gamma+\gamma^{^{\prime}%
})\mid^{2}=O(\rho^{-2k\alpha_{1}}),
\end{equation}%
\begin{equation}
\Psi_{N}=b(N,\gamma)e^{i(\gamma+t,x)}+\sum_{\gamma^{^{\prime}}\in
\Gamma((k-1)\rho^{\alpha})}b(N,\gamma+\gamma^{^{\prime}})e^{i(\gamma
+t+\gamma^{^{\prime}},x)}+H_{k}(x),
\end{equation}
where $\Gamma^{c}(j)\equiv\Gamma\backslash(\Gamma(j\rho^{\alpha})\cup\{0\})$
and $\parallel H_{k}(x)\parallel=O(\rho^{-k\alpha_{1}}).$ The case $k=1$ is
proved due to (14). Assume that (56) is true for $k=j$ . Then using (57) for
$\ \ \ k=j,$ and (3), we have $\Psi_{N}(x)(q(x))=H(x)+O(\rho^{-j\alpha_{1}}),$
where $H(x)$ is a linear combination of $e^{i(\gamma+t+\gamma^{^{\prime}},x)}$
for $\gamma^{^{\prime}}\in\Gamma(j\rho^{\alpha})\cup\{0\}.$ Hence
$(H(x),e^{i(\gamma+t+\gamma^{^{\prime}},x)})=0$ for $\gamma^{^{\prime}}%
\in\Gamma^{c}(j).$ So using (16) and (50), we get%
\begin{equation}
\sum_{\gamma^{^{\prime}}}\mid b(N,\gamma+\gamma^{^{\prime}})\mid^{2}%
=\sum_{\gamma^{^{\prime}}}\mid\dfrac{(O(\rho^{-j\alpha_{1}}),e^{i(\gamma
+t+\gamma^{^{\prime}},x)})}{\Lambda_{N}-\mid\gamma+\gamma^{^{\prime}}%
+t\mid^{2l}}\mid^{2}=O(\rho^{-2(j+1)\alpha_{1}}),
\end{equation}
where the summation is taken under conditions $\gamma^{^{\prime}}\in\Gamma
^{c}(j)$, $\gamma+\gamma^{^{\prime}}\notin K.$ On the other hand, using
$\alpha_{1}=3\alpha,$ (108), and the definition of $n$, we obtain
\[
\sum_{\gamma^{^{\prime}}\in K\backslash\{\gamma\}}\mid b(N,\gamma^{^{\prime}%
})\mid^{2}=O(\rho^{-2n\alpha_{1}}).
\]
This with (58) implies (56) for $k=j+1.$ Thus (57) is also proved. Here
$b(N,\gamma)$ and $b(N,\gamma+\gamma^{^{\prime}})$ for $\gamma^{^{\prime}}%
\in\Gamma((n-1)\rho^{\alpha})$ can be calculated as follows. First we express
$b(N,\gamma+\gamma^{^{\prime}})$ by $b(N,\gamma)$. For this we apply (18) for
$b(N,\gamma+\gamma^{^{\prime}}),$ where $\gamma^{^{\prime}}\in\Gamma
((n-1)\rho^{\alpha}),$ that is, in (18) replace $\gamma^{^{\prime}}$ by
$\gamma+\gamma^{^{\prime}}$. Iterate it $n$ times and every times isolate the
terms with multiplicand $b(N,\gamma).$ In other word apply (18) for
$b(N,\gamma+\gamma^{^{\prime}})$ and isolate the terms with multiplicand
$b(N,\gamma).$ Then apply (18) for $b(N,\gamma+\gamma^{^{\prime}}-\gamma_{1})$
when $\gamma^{^{\prime}}-\gamma_{1}\neq0.$ Then apply (18) for

$b(N,\gamma+\gamma^{^{\prime}}-\sum_{i=1}^{2}\gamma_{i})$ when $\gamma
^{^{\prime}}-\sum_{i=1}^{2}\gamma_{i}\neq0,$ etc. Apply (18) for

$b(N,\gamma+\gamma^{^{\prime}}-\sum_{i=1}^{j}\gamma_{i})$ when $\gamma
^{^{\prime}}-\sum_{i=1}^{j}\gamma_{i}\neq0,$ where $\gamma_{i}\in\Gamma
(\rho^{\alpha}),$

$j=3,4,...,n-1.$ Then using (4) and the relations

$\mid\Lambda_{N}-\mid\gamma+t+\gamma^{^{\prime}}-\sum_{i=1}^{j}\gamma_{i}%
\mid^{2l}\mid>\frac{1}{2}\rho^{\alpha_{1}}$ ( see (20) and take into account that

$\gamma^{^{\prime}}-\sum_{i=1}^{j}\gamma_{i}\in\Gamma(p\rho^{\alpha}),$ since
$p>2n$), $\Lambda_{N}=P(\gamma+t)+O(\rho^{-n\alpha_{1}}),$ where
$P(\gamma+t)=\mid\gamma+t\mid^{2l}+F_{[\frac{p}{3}]}(\gamma+t)$ ( see Theorem
3), we obtain%
\begin{equation}
b(N,\gamma+\gamma^{^{\prime}})=\sum_{k=1}^{n-1}A_{k}(\gamma^{^{\prime}%
})b(N,\gamma)+O(\rho^{-n\alpha_{1}}),
\end{equation}
where

$A_{1}(\gamma^{^{\prime}})=\dfrac{q_{\gamma^{^{\prime}}}}{P(\gamma
+t)-\mid\gamma+\gamma^{^{\prime}}+t\mid^{2l}}=\dfrac{q_{\gamma^{^{\prime}}}%
}{\mid\gamma+t\mid^{2l}-\mid\gamma+\gamma^{^{\prime}}+t\mid^{2l}}+O(\frac
{1}{\rho^{3\alpha_{1}}}),$%

\[
A_{k}(\gamma^{^{\prime}})=%
%TCIMACRO{\dsum _{\gamma_{1},...,\gamma_{k-1}}}%
%BeginExpansion
{\displaystyle\sum_{\gamma_{1},...,\gamma_{k-1}}}
%EndExpansion
\dfrac{q_{\gamma_{1}}q_{\gamma_{2}}...q_{\gamma_{k-1}}q_{\gamma^{^{\prime}%
}-\gamma_{1}-\gamma_{2}-...-\gamma_{k-1}}}{\prod_{j=0}^{k-1}(P(\gamma
+t)-\mid\gamma+t+\gamma^{^{\prime}}-\sum_{i=1}^{j}\gamma_{i}\mid^{2l})}%
=O(\rho^{-k\alpha_{1}}),
\]%
\begin{equation}
\sum_{\gamma^{\ast}\in\Gamma((n-1)\rho^{\alpha})}\mid A_{1}(\gamma^{\ast}%
)\mid^{2}=O(\rho^{-2\alpha_{1}}),\sum_{\gamma^{\ast}\in\Gamma((n-1)\rho
^{\alpha})}\mid A_{k}(\gamma^{\ast})\mid=O(\rho^{-k\alpha_{1}})
\end{equation}
for $k>1.$ Now from (57) for $k=n$ and (59), we obtain
\begin{align*}
\Psi_{N}(x)  &  =b(N,\gamma)e^{i(\gamma+t,x)}+\\
&  \sum_{\gamma^{\ast}\in\Gamma((n-1)\rho^{\alpha})}\sum_{k=1}^{n-1}%
(A_{k}(\gamma^{\ast})b(N,\gamma)+O(\rho^{-n\alpha_{1}}))e^{i(\gamma
+t+\gamma^{\ast},x)})+H_{n}(x).
\end{align*}
Using the equalities $\parallel\Psi_{N}\parallel=1,$ $\arg b(N,\gamma)=0,$
$\parallel H_{n}\parallel=O(\rho^{-n\alpha_{1}})$ and taking into account that
the functions $e^{i(\gamma+t,x)},$ $H_{n}(x),$ $e^{i(\gamma+t+\gamma^{\ast
},x)},$ $(\gamma^{\ast}\in\Gamma((n-1)\rho^{\alpha}))$ are orthogonal, we get

$1=\mid b(N,\gamma)\mid^{2}+\sum_{k=1}^{n-1}(\sum_{\gamma^{\ast}\in
\Gamma((n-1)\rho^{\alpha})}\mid A_{k}(\gamma^{\ast})b(N,\gamma)\mid^{2}%
+O(\rho^{-n\alpha_{1}})),$
\begin{equation}
b(N,\gamma)=(1+\sum_{k=1}^{n-1}(\sum_{\gamma^{\ast}\in\Gamma((n-1)\rho
^{\alpha})}\mid A_{k}(\gamma^{\ast})\mid^{2}))^{-\frac{1}{2}}+O(\rho
^{-n\alpha_{1}}))
\end{equation}
(see the second equality in (60)). Thus from (59), we obtain
\begin{equation}
b(N,\gamma+\gamma^{^{\prime}})=(\sum_{k=1}^{n-1}A_{k}(\gamma^{^{\prime}%
}))(1+\sum_{k=1}^{n-1}\sum_{\gamma^{\ast}}\mid A_{k}(\gamma^{\ast})\mid
^{2})^{-\frac{1}{2}}+O(\rho^{-n\alpha_{1}}).
\end{equation}
Consider the case $n=2.$ By (61), (60), (62) we have $b(N,\gamma
)=1+O(\rho^{-2\alpha_{1}}),$

$b(N,\gamma+\gamma^{^{\prime}})=A_{1}(\gamma^{^{\prime}})+O(\rho^{-2\alpha
_{1}})=\dfrac{q_{\gamma^{^{\prime}}}}{\mid\gamma+t\mid^{2l}-\mid\gamma
+\gamma^{^{\prime}}+t\mid^{2l}}+O(\rho^{-2\alpha_{1}})$ for all $\gamma
^{^{\prime}}\in\Gamma(\rho^{\alpha}).$ These and (57) for $k=2$ imply the
formula for $\Phi_{1}$
\end{proof}

\section{Simple Sets and Bethe-Sommerfeld conjecture}

In this section we construct a part of the simple set $B$ in the neighbourhood
of the surface $S_{\rho}\equiv\{x\in U^{1}(2\rho^{\alpha_{1}},p):F(x)=\rho
^{2l}\},$ where $F(x)=\mid x\mid^{2l}+F_{k_{1}-1}(x)$ for $x=\gamma+t$ \ is
defined in (49) and in introduction ( see step 1). Due to (49) it is natural
to call $S_{\rho}$ the approximated isoenergetic surfaces in the non-resonance
domain. As we noted in introduction ( see the inequality (12)) the
non-resonance eigenvalue $\Lambda(\gamma+t),$ where $\Lambda(\gamma
+t)=\Lambda_{N(\gamma+t)}(t)$ is defined in Remark 3, does not coincide with
other non-resonance eigenvalue $\Lambda(\gamma+t+b)$ if $\mid F(\gamma
+t)-F(\gamma+t+b)\mid>2\varepsilon_{1}$ for $\gamma+t+b\in U^{1}(\rho
^{\alpha_{1}},p)$ and $b\in\Gamma\backslash\{0\}$ . Therefore we eliminate
\begin{equation}
P_{b}\equiv\{x:x,x+b\in U^{1}(\rho^{\alpha_{1}},p),\mid F(x)-F(x+b)\mid
<3\varepsilon_{1}\}
\end{equation}
for $b\in\Gamma\backslash\{0\}$ from $S_{\rho}$. Denote the remaining part of
$S_{\rho}$ by $S_{\rho}^{^{\prime}}.$ Then we consider the $\varepsilon$ neighbourhood

$U_{\varepsilon}(S_{\rho}^{^{\prime}})=\cup_{a\in S_{\rho}^{^{\prime}}%
}U_{\varepsilon}(a)\}$ of $S_{\rho}^{^{\prime}}$, where $U_{\varepsilon
}(a)=\{x\in\mathbb{R}^{d}:\mid x-a\mid<\varepsilon\},$

and prove that in this set the first simplicity condition (12) holds (see
Lemma 2(a)). Denote by $Tr(E)\equiv\{\gamma+x\in U_{\varepsilon}(S_{\rho
}^{^{\prime}}):\gamma\in\Gamma,x\in E\}$ and

$Tr_{F^{\star}}(E)\equiv\{\gamma+x\in F^{\star}:\gamma\in\Gamma,x\in E\}$ the
translations of $E\subset\mathbb{R}^{d}$ into $U_{\varepsilon}(S_{\rho
}^{^{\prime}})$ and $F^{\star}$ respectively. In order that the second
simplicity condition (13) holds, we discard from $U_{\varepsilon}(S_{\rho
}^{^{\prime}})$ the translation $Tr(A(\rho))$ of
\begin{equation}
A(\rho)\equiv\cup_{k=1}^{d-1}(\cup_{\gamma_{1},\gamma_{2},...,\gamma_{k}%
\in\Gamma(p\rho^{\alpha})}(\cup_{i=1}^{b_{k}}A_{k,i}(\gamma_{1},\gamma
_{2},...,\gamma_{k}))),
\end{equation}

where $A_{k,i}(\gamma_{1},...,\gamma_{k})=$

$\{x\in(\cap_{i=1}^{k}V_{\gamma_{i}}^{1}(\rho^{\alpha_{k}})\backslash
E_{k+1}^{1})\cap K_{\rho}:\lambda_{i}(x)\in(\rho^{2l}-3\varepsilon_{1}%
,\rho^{2l}+3\varepsilon_{1})\},$

$\lambda_{i}(x),$ $b_{k}$ is defined in Theorem 2 and
\begin{equation}
K_{\rho}=\{x\in\mathbb{R}^{d}:\mid\mid x\mid^{2l}-\rho^{2l}\mid<\rho
^{\alpha_{1}}\}.
\end{equation}
As a result, we construct the part $U_{\varepsilon}(S_{\rho}^{^{\prime}%
})\backslash Tr(A(\rho))$ of the simple set $B$ (see Theorem 5(a)), which
contains the intervals $\{a+sb:s\in\lbrack-1,1]\}$ such that $\Lambda
(a-b)<\rho^{2l},$ $\Lambda(a+b)>\rho^{2l}$\ and $\Lambda(\gamma+t)$ \ is
continuous on this intervals. Hence there exists $\gamma+t$ such that
$\Lambda(\gamma+t)=\rho^{2l}.$ It implies the validity of Bethe-Sommerfeld
conjecture for $L(l,q)$.

\begin{lemma}
$(a)$ If $x\in U_{\varepsilon}(S_{\rho}^{^{\prime}})$ and $x+b\in U^{1}%
(\rho^{\alpha_{1}},p),$ where $b\in\Gamma,$ then
\begin{equation}
\mid F(x)-F(x+b)\mid>2\varepsilon_{1},
\end{equation}
where $\varepsilon=\frac{\varepsilon_{1}}{7l\rho^{2l-1}},\varepsilon_{1}%
=\rho^{-d-2\alpha},$ hence for $\gamma+t\in U_{\varepsilon}(S_{\rho}%
^{^{\prime}})$ the first simplicity condition (12) holds.

$(b)$ If $x\in U_{\varepsilon}(S_{\rho}^{^{\prime}}),$ then $x+b\notin
U_{\varepsilon}(S_{\rho}^{^{\prime}})$ for all $b\in\Gamma$ $.$

$(c)$If $E\subset\mathbb{R}^{d}$ is bounded set, then $\mu(Tr(E))\leq\mu(E)$.

$(d)$ If $E\subset U_{\varepsilon}(S_{\rho}^{^{\prime}}),$ then $\mu
(Tr_{F^{\star}}(E))=\mu(E).$
\end{lemma}

\begin{proof}
$(a)$ If $x\in U_{\varepsilon}(S_{\rho}^{^{\prime}}),$ then there exists a
point $a$ in $S_{\rho}^{^{\prime}}$ such that $x\in U_{\varepsilon}(a)$. Since
$S_{\rho}^{^{\prime}}\cap P_{b}=\emptyset$ ( see (63) and def. of $S_{\rho
}^{^{\prime}}$), we have
\begin{equation}
\mid F(a)-F(a+b)\mid\geq3\varepsilon_{1}%
\end{equation}
On the other hand, using (47) and the obvious relations

$\mid x\mid<\rho+1,$ $\mid x-a\mid<\varepsilon,$ $\mid x+b-a-b\mid
<\varepsilon,$ we obtain
\begin{equation}
\mid F(x)-F(a)\mid<3l\rho^{2l-1}\varepsilon,\mid F(x+b)-F(a+b)\mid
<3l\rho^{2l-1}\varepsilon
\end{equation}
These inequalities together with (67) give (66), since $6l\rho^{2l-1}%
\varepsilon<\varepsilon_{1}.$

$(b)$ If $x$ and $x+b$ lie in $U_{\varepsilon}(S_{\rho}^{^{\prime}}),$ then
there exist points $a$ and $c$ in $S_{\rho}^{^{\prime}}$ such that $x\in
U_{\varepsilon}(a)$ and $x+b\in U_{\varepsilon}(c).$ Repeating the proof of
(68), we get

$\mid F(c)-F(x+b)\mid<3l\rho^{2l-1}\varepsilon.$ This, the first inequality in
(68), and the relations $F(a)=\rho^{2l},F(c)=\rho^{2l}$ (see the definition of
$S_{\rho})$ give

$\mid F(x)-F(x+b)\mid<\varepsilon_{1},$ which contradicts (66).

$(c)$ Clearly, for any bounded set $E$ there are only finite number of the
vectors $\gamma_{1},\gamma_{2},...,\gamma_{s}$ such that $E(k)\equiv
(E+\gamma_{k})\cap U_{\varepsilon}(S_{\rho}^{^{\prime}})\neq\emptyset$ for
$k=1,2,...,s$ and $Tr(E)$ is the union of the sets $E(k)$. For $E(k)-\gamma
_{k}$ we have the relations $\mu(E(k)-\gamma_{k})=\mu(E(k)),$ $E(k)-\gamma
_{k}\subset E.$ Moreover, by $(b)$

$(E(k)-\gamma_{k})\cap(E(j)-\gamma_{j})=\emptyset$ for $k\neq j.$ Therefore
$(c)$ is true.

$(d)$ Now let $E\subset U_{\varepsilon}(S_{\rho}^{^{\prime}}).$ Then by $(b)$
the set $E$ can be divided into finite number of the pairwise disjoint sets
$E_{1},E_{2},...,E_{n}$ such that there exist the vectors $\gamma_{1}%
,\gamma_{2},...,\gamma_{n}$ satisfying $(E_{k}+\gamma_{k})\subset F^{\star},$
$(E_{k}+\gamma_{k})\cap(E_{j}+\gamma_{j})\neq\emptyset$ for $k,j=1,2,...,n$
and $k\neq j.$ Using $\mu(E_{k}+\gamma_{k})=\mu(E_{k}),$ we get the proof of
$(d),$ because $Tr_{F^{\star}}(E)$ and $E$ are union of the pairwise disjoint
sets $E_{k}+\gamma_{k}$ and $E_{k}$ for $k=1,2,...,n$ respectively
\end{proof}

\begin{theorem}
$(a)$ The set $U_{\varepsilon}(S_{\rho}^{^{\prime}})\backslash Tr(A(\rho))$ is
a subset of $B,$ hence if $\gamma+t$ lies in this subset, then Theorem 3 and
Theorem 4 hold. For every connected open subset $E$ of $U_{\varepsilon
}(S_{\rho}^{^{\prime}})\backslash Tr(A(\rho)$ there exists a unique index $N$
such that

$\Lambda(\gamma+t)=\Lambda_{N}(t)$ for $\gamma+t\in E,$ where $\Lambda
(\gamma+t)$ is defined in Remark 3.

$(b)$ For the part $V_{\rho}\equiv S_{\rho}^{^{\prime}}\backslash
U_{\varepsilon}(Tr(A(\rho)))$ of the approximated isoenergetic surface
$S_{\rho}$ the following holds
\begin{equation}
\mu(V_{\rho})>(1-c_{9}\rho^{-\alpha}))\mu(B(\rho)).
\end{equation}
Moreover, $U_{\varepsilon}(V_{\rho})$ lies in the subset $U_{\varepsilon
}(S_{\rho}^{^{\prime}})\backslash Tr(A(\rho))$ of the simple set $B.$

$(c)$ The number $\rho^{2l}$ for $\rho\gg1$ lies in the spectrum of
$L(l,q(x)),$ that is, the number of the gaps in the spectrum of $L(l,q(x))$ is
finite, where $l\geq1,$ $q(x)\in W_{2}^{s_{0}}(\mathbb{R}^{d}/\Omega),$
$d\geq2,$ $s_{0}=\frac{3d-1}{2}(3^{d}+d+2)+\frac{1}{4}d3^{d}+d+6,$ and
$\Omega$ is an arbitrary lattice.
\end{theorem}

\begin{proof}
$(a)$ To prove that $U_{\varepsilon}(S_{\rho}^{^{\prime}})\backslash
Tr(A(\rho))\subset B$ we need to show that for each point $\gamma+t$ of
$U_{\varepsilon}(S_{\rho}^{^{\prime}})\backslash Tr(A(\rho))$ the simplicity
conditions (12), (13) hold and $U_{\varepsilon}(S_{\rho}^{^{\prime}%
})\backslash Tr(A(\rho))\subset U^{1}(\rho^{\alpha_{1}},p).$ By lemma 2(a),
the condition (12) holds. Now we prove that (13) holds too. Since $\gamma+t\in
U_{\varepsilon}(S_{\rho}^{^{\prime}}),$ there exists $a\in S_{\rho}^{^{\prime
}}$ such that $\gamma+t\in U_{\varepsilon}(a).$ The first inequality in (68)
and equality $F(a)=\rho^{2l}$ imply
\begin{equation}
F(\gamma+t)\in(\rho^{2l}-\varepsilon_{1},\rho^{2l}+\varepsilon_{1})
\end{equation}
for $\gamma+t\in U_{\varepsilon}(S_{\rho}^{^{\prime}}).$ On the other hand
$\gamma+t\notin Tr(A(\rho)).$ It means that for any $\gamma^{^{\prime}}%
\in\Gamma,$ we have $\gamma^{^{\prime}}+t\notin A(\rho).$ If $\gamma
^{^{\prime}}\in K$ and $\gamma^{^{\prime}}+t\in E_{k}^{1}\backslash
E_{k+1}^{1},$ then by definition of $K$ ( see introduction) the inequality
$\mid F(\gamma+t)-\mid\gamma^{^{\prime}}+t\mid^{2l}\mid<\frac{1}{3}%
\rho^{\alpha_{1}}$ holds. This and (70) imply that $\gamma^{^{\prime}}%
+t\in(E_{k}^{1}\backslash E_{k+1}^{1})\cap K_{\rho}$ ( see (65) for the
definition of $K_{\rho}$). Since $\gamma^{^{\prime}}+t\notin A(\rho),$ we have
$\lambda_{i}(\gamma^{^{\prime}}+t)\notin(\rho^{2l}-3\varepsilon_{1},\rho
^{2l}+3\varepsilon_{1})$ for $\gamma^{^{\prime}}\in K$ and $\gamma^{^{\prime}%
}+t\in E_{k}^{1}\backslash E_{k+1}^{1}.$ Therefore (13) follows from (70).
Moreover, it is clear that the inclusion $S_{\rho}^{^{\prime}}\subset
U^{1}(2\rho^{\alpha_{1}},p)$ ( see definition of $S_{\rho}$ and $S_{\rho
}^{^{\prime}}$) implies that $U_{\varepsilon}(S_{\rho}^{^{\prime}})\subset
U^{1}(\rho^{\alpha_{1}},p).$ Thus $U_{\varepsilon}(S_{\rho}^{^{\prime}%
})\backslash Tr(A(\rho))\subset B.$

Now let $E$ be a connected open subset of $U_{\varepsilon}(S_{\rho}^{^{\prime
}})\backslash Tr(A(\rho)\subset B.$ By Theorem 3 and Remark 3 for $a\in
E\subset U_{\varepsilon}(S_{\rho}^{^{\prime}})\backslash Tr(A(\rho)$ there
exists a unique index $N(a)$ such that $\Lambda(a)=\Lambda_{N(a)}(a)$,
$\Psi_{a}(x)=\Psi_{N(a),a}(x)$, $\mid(\Psi_{N(a),a}(x),e^{i(a,x)})\mid
^{2}>\frac{1}{2}$ and $\Lambda(a)$ is a simple eigenvalue. On the other hand,
for fixed $N$ the functions $\Lambda_{N}(t)$ and $(\Psi_{N,t}(x),e^{i(t,x)})$
are continuous in a neighborhood of $a$ if $\Lambda_{N}(a)$ is a simple
eigenvalue. Therefore for each $a\in E$ there exists a neighborhood
$U(a)\subset E$ of $a$ such that $\mid(\Psi_{N(a),y}(x),e^{i(y,x)})\mid
^{2}>\frac{1}{2}$, for $y\in U(a).$ Since for $y\in E$ there is a unique
integer $N(y)$ satisfying $\mid(\Psi_{N(y),y}(x),e^{i(y,x)})\mid^{2}>\frac
{1}{2},$ we have $N(y)=N(a)$ for $y\in U(a).$ Hence we proved that%
\begin{equation}
\forall a\in E,\exists U(a)\subset E:N(y)=N(a),\forall y\in U(a).
\end{equation}

Now let $a_{1}$ and $a_{2}$ be two points of $E$ , and let $C\subset E$ be the
arc that joins these points. Let $U(y_{1}),U(y_{2}),...,U(y_{k})$ be a finite
subcover of the open cover $\cup_{a\in C}U(a)$ of the compact $C,$ where
$U(a)$ is the neighborhood of $a$ satisfying (71). By (71), we have
$N(y)=N(y_{i})=N_{i}$ for $y\in U(y_{i}).$ Clearly, if

$U(y_{i})\cap U(y_{j})\neq\emptyset,$ then $N_{i}=N(z)=N_{j},$ where $z\in
U(y_{i})\cap U(y_{j})$. Thus $N_{1}=N_{2}=...=N_{k}$ and $N(a_{1})=N(a_{2}).$

$(b)$ To prove the inclusion $U_{\varepsilon}(V_{\rho})\subset$
$U_{\varepsilon}(S_{\rho}^{^{\prime}})\backslash Tr(A(\rho))$ we need to show
that if $a\in V_{\rho},$ then $U_{\varepsilon}(a)\subset U_{\varepsilon
}(S_{\rho}^{^{\prime}})\backslash Tr(A(\rho)).$ This is clear, since the
relations $a\in V_{\rho}\subset S_{\rho}^{^{\prime}}$ imply that
$U_{\varepsilon}(a)\subset U_{\varepsilon}(S_{\rho}^{^{\prime}})$ and the
relation $a\notin U_{\varepsilon}(Tr(A(\rho)))$ implies that $U_{\varepsilon
}(a)\cap Tr(A(\rho))=\emptyset.$ To prove (69) first we estimate the measure
of $S_{\rho},S_{\rho}^{^{\prime}},U_{2\varepsilon}(A(\rho))$, namely we prove
\begin{align}
\mu(S_{\rho})  &  >(1-c_{10}\rho^{-\alpha})\mu(B(\rho)),\\
\mu(S_{\rho}^{^{\prime}})  &  >(1-c_{11}\rho^{-\alpha})\mu(B(\rho)),\\
\mu(U_{2\varepsilon}(A(\rho)))  &  =O(\rho^{-\alpha})\mu(B(\rho))\varepsilon
\end{align}
( see below, Estimations 1, 2, 3). The estimation (69) of the measure of the
set $V_{\rho}$ is done in Estimation 4 by using Estimations 1, 2, 3.

$(c)$ Since $F(a)=\rho^{2l}$ for $a=(a_{1},a_{2},...,a_{d})=\sum_{i=1}%
^{d}a_{i}e_{i}\in V_{\rho}\subset S_{\rho},$ it follows from (24) that
$\rho-1<\mid a\mid<\rho+1,$ and there exists an index $i$ such that $\mid
a_{i}\mid>\frac{1}{d}\rho$. Without loss of generality it can be assumed that
$\ a_{i}>0.$ Then (47) and (49) imply that $F(a-\varepsilon e_{i})<\rho
^{2l}-c_{11}\varepsilon_{1},$ $F(a+\varepsilon e_{i})>\rho^{2l}+c_{11}%
\varepsilon_{1}$ and $\Lambda(a-\varepsilon e_{i})<\rho^{2l},$ $\Lambda
(a+\varepsilon e_{i})>\rho^{2l}.$ Since $\Lambda(\gamma+t)$ \ is continuous in
$U_{\varepsilon}(a)\subset U_{\varepsilon}(S_{\rho}^{^{\prime}})\backslash
Tr(A(\rho)$ ( see Theorem 5(a)), there exists $y(a,i)\in(a-\varepsilon
e_{i},a+\varepsilon e_{i})$ such that $\Lambda(y(a,i))=\rho^{2l}.$ The Theorem
is proved
\end{proof}

In Estimations 1-4 we use the notations: $G(+i,a)=\{x\in G,x_{i}>a\},$
$G(-i,a)=\{x\in G,x_{i}<-a\},$ where $x=(x_{1},x_{2},...,x_{d}),a>0.$ It is
not hard to verify that for any subset $G$ of $U_{\varepsilon}(S_{\rho
}^{^{\prime}})\cup U_{2\varepsilon}(A(\rho))$ , that is, for all considered
sets $G$ in these estimations, and for any $x\in G$ the followings hold
\begin{equation}
\rho-1<\mid x\mid<\rho+1,\text{ }G\subset(\cup_{i=1}^{d}(G(+i,\rho d^{-1})\cup
G(-i,\rho d^{-1}))
\end{equation}
Indeed, if $x\in S_{\rho}^{^{\prime}},$ then $F(x)=\rho^{2l}$ and by (24) we
have $\mid x\mid=\rho+O(\rho^{-1-\alpha_{1}}).$ Hence the inequalities in (75)
hold for $x\in U_{\varepsilon}(S_{\rho}^{^{\prime}}).$ If $x\in$ $A(\rho),$
then by definition of $A(\rho),$ we have $x\in K_{\rho},$ and hence $\mid
x\mid=\rho+O(\rho^{-1+\alpha_{1}})$. Therefore the inequalities in (75) hold
for $x\in U_{2\varepsilon}(A(\rho))$ too. The inclusion in (75) follows from
these inequalities.

If $G$ $\subset S_{\rho},$ then by (47) we have $\frac{\partial F(x)}{\partial
x_{k}}>0$ for $x\in G(+k,\rho^{-\alpha})$. Therefore to calculate the measure
of $G(+k,a)$ for $a\geq\rho^{-\alpha}$ we use the formula
\begin{equation}
\mu(G(+k,a))=\int\limits_{\Pr_{k}(G(+k,a))}(\frac{\partial F}{\partial x_{k}%
})^{-1}\mid grad(F)\mid dx_{1}...dx_{k-1}dx_{k+1}...dx_{d},
\end{equation}
where $\Pr_{k}(G)\equiv\{(x_{1},x_{2},...,x_{k-1},x_{k+1},x_{k+2}%
,...,x_{d}):x\in G\}$ is the projection of $G$ on the hyperplane $x_{k}=0.$
Instead of $\Pr_{k}(G)$ we write $\Pr(G)$ if $k$ is unambiguous. If $D$ is
$s-$dimensional subset of $\mathbb{R}^{s},$ then to estimate $\mu(D),$ we use
the formula
\begin{equation}
\mu(D)=\int\limits_{\Pr_{k}(D)}\mu(D(x_{1},...x_{k-1},x_{k+1},...,x_{s}%
))dx_{1}...dx_{k-1}dx_{k+1}...dx_{s},
\end{equation}
where $D(x_{1},...x_{k-1},x_{k+1},...,x_{s})=\{x_{k}:(x_{1},x_{2}%
,...,x_{s})\in D\}.$

ESTIMATION 1. Here we prove (72) by using (76). During this estimation the set
$S_{\rho}$ is redenoted by $G.$ First we estimate $\mu(G(+1,a))$ for
$a=\rho^{1-\alpha}$ by using (76) for $k=1$ and the following relations
\begin{equation}
\frac{\partial F(x)}{\partial x_{1}}=l\mid x\mid^{2(l-1)}(2x_{1}%
+O(\rho^{-2\alpha}))
\end{equation}%
\begin{equation}
\frac{\partial F}{\partial x_{1}}>\rho^{1-\alpha},\text{ }(\frac{\partial
F}{\partial x_{1}})^{-1}\mid grad(F)\mid=\frac{\rho}{\sqrt{\rho^{2}-x_{2}%
^{2}-x_{3}^{2}-...-x_{d}^{2}}}+O(\rho^{-\alpha}),
\end{equation}%
\begin{equation}
\Pr(G(+1,a))\supset\Pr(A(+1,2a)),
\end{equation}
where $x\in G(+1,a),$ $A=B(\rho)\cap U^{1}(3\rho^{\alpha_{1}},p),$
$B(\rho)=\{x\in\mathbb{R}^{d}:\mid x\mid=\rho\}.$ The estimations (78) and
(79) follow from (47). Now we prove (80). If

$(x_{2},...,x_{d})\in\Pr_{1}(A(+1,2a)),$ then there exists $x_{1}$ such that
\begin{equation}
x_{1}>2a=2\rho^{1-\alpha},\text{ }x_{1}^{2}+x_{2}^{2}+...+x_{d}^{2}=\rho
^{2},\mid\sum_{i\geq1}(2x_{i}b_{i}-b_{i}^{2})\mid\geq3\rho^{\alpha_{1}}%
\end{equation}
for all $(b_{1},b_{2},...,b_{d})\in\Gamma(p\rho^{\alpha}).$ Therefore for
$h=\rho^{-\alpha},$ we have

$(x_{1}+h)^{2}+x_{2}^{2}+...+x_{q}^{2}>\rho^{2}+\rho^{-\alpha},(x_{1}%
-h)^{2}+x_{2}^{2}+...+x_{q}^{2}<\rho^{2}-\rho^{-\alpha}.$ These inequalities
and (24) give

$F(x_{1}+h,x_{2},...,x_{d})>\rho^{2l}$, $F(x_{1}-h,x_{2},...,x_{d})<\rho^{2l}%
$. Since $F$ is a continuous function, there exists $y_{1}\in(x_{1}%
-h,x_{1}+h)$ such that
\begin{equation}
y_{1}>a,F(y_{1},x_{2},...,x_{d})=\rho^{2l},\text{ }\mid2y_{1}b_{1}-b_{1}%
^{2}+\sum_{i\geq2}(2x_{i}b_{i}-b_{i}^{2})\mid>2\rho^{\alpha_{1}},
\end{equation}
because the expression under the absolute value in (82) differ from the
expression under the absolute value in (81) by $2(y_{1}-x_{1})b_{1},$ where
$\mid y_{1}-x_{1}\mid<h=\rho^{\alpha},$ $b_{1}<p\rho^{\alpha},$ $\mid
2(y_{1}-x_{1})b_{1}\mid<2p\rho^{2\alpha}<$ $\rho^{\alpha_{1}}.$ The relations
in (82) means that $(x_{2},...,x_{d})\in\Pr G(+1,a).$ Hence (80) is proved.
Now (76), (79), and the obvious relation $\mu(\Pr G(+1,a))=O(\rho^{d-1})$ (
see the inequalities in (75)) imply that%
\[
\mu(G(+1,a))=\int\limits_{\Pr(G(+1,a))}\frac{\rho}{\sqrt{\rho^{2}-x_{2}%
^{2}-x_{3}^{2}-...-x_{d}^{2}}}dx_{2}dx_{3}...dx_{d}+O(\frac{1}{\rho^{\alpha}%
})\mu(B(\rho))
\]%
\[
\geq\int\limits_{\Pr(A(+1,2a))}\frac{\rho}{\sqrt{\rho^{2}-x_{2}^{2}-x_{3}%
^{2}-...-x_{d}^{2}}}dx_{2}dx_{3}...dx_{d}-c_{12}\rho^{-\alpha}\mu(B(\rho))
\]%
\[
=\mu(A(+1,2a))-c_{12}\rho^{-\alpha}\mu(B(\rho)).
\]
Similarly, $\mu(G(-1,a))\geq\mu(A(-1,2a))-c_{12}\rho^{-\alpha}\mu(B(\rho)).$
Now using \ the inequality $\mu(G)\geq\mu(G(+1,a))+\mu(G(-1,a))$ we get

$\mu(G)\geq\mu(A(-1,2a))+\mu(A(+1,2a))-2c_{12}\rho^{-\alpha}\mu(B(\rho)).$ On
the other hand it follows from the obvious relation\ 

\ $\mu(\{x\in B(\rho):-2a\leq x_{1}\leq2a\})=O(\rho^{-\alpha})\mu(B(\rho))$ that

$\mu(A(-1,2a))+\mu(A(+1,2a))\geq\mu(A)-c_{13}\rho^{-\alpha}\mu(B(\rho)).$ Therefore

$\mu(G)>\mu(A)-c_{14}\rho^{-\alpha}\mu(B(\rho)).$ It implies (72), since

$\mu(A))=(1+O(\rho^{-\alpha}))\mu(B(\rho))$ (see (45) ).

ESTIMATION 2 Here we prove (73). For this we estimate the measure of the set
$S_{\rho}\cap P_{b}$ ( see (63)) by using (76). During this estimation the set
$S_{\rho}\cap P_{b}$ is redenoted by $G$. We choose the coordinate axis so
that the direction of $b$ coincides with the direction of $(1,0,0,...,0),$
i.e., $b=(b_{1},0,0,...,0)$ and $b_{1}>0$. It follows from the definitions of
$S_{\rho},P_{b}$ and $F(x)$ ( see the beginning of this section and (63))
that, if $(x_{1},x_{2},...,x_{d})\in G=$ $S_{\rho}\cap P_{b},$ then
\begin{align}
(x_{1}^{2}+x_{2}^{2}+...+x_{d}^{2})^{l}+F_{k_{1}-1}(x)  &  =\rho^{2l},\\
((x_{1}-b_{1})^{2}+x_{2}^{2}+x_{3}^{2}+...+x_{d}^{2})^{l}+F_{k_{1}-1}(x+b)  &
=\rho^{2l}+h,
\end{align}
where $h\in(-3\varepsilon_{1},3\varepsilon_{1}),$ and by (24), it follows
that
\begin{align}
(x_{1}^{2}+x_{2}^{2}+x_{3}^{2}+...+x_{d}^{2})  &  =\rho^{2}+O(\rho
^{-\alpha_{1}})\\
((x_{1}-b_{1})^{2}+x_{2}^{2}+x_{3}^{2}+...+x_{d}^{2})  &  =\rho^{2}%
+O(\rho^{-\alpha_{1}}).
\end{align}
Subtracting (85) from (86), we get%
\begin{equation}
(2x_{1}-b_{1})b_{1}=O(\rho^{-\alpha_{1}}).
\end{equation}
Now (87) and the inequalities in (75) imply
\begin{equation}
\mid b_{1}\mid<2\rho+3,\text{ }x_{1}=\frac{b_{1}}{2}+O(\rho^{-\alpha_{1}}%
b_{1}^{-1}),\mid x_{1}^{2}-(\frac{b_{1}}{2})^{2}\mid=O(\rho^{-\alpha_{1}})
\end{equation}
Consider two cases. Case 1: $b\in\Gamma_{1},$ where $\Gamma_{1}=\{b\in
\Gamma:\mid\rho^{2}-(\frac{b_{1}}{2})^{2}\mid<3d\rho^{-2\alpha}\}.$ In this
case using $\alpha_{1}=3\alpha,$ the last equality in (88), and (85), we
obtain
\begin{equation}
x_{1}^{2}=\rho^{2}+O(\rho^{-2\alpha}),\mid x_{1}\mid=\rho+O(\rho^{-2\alpha
-1}),\text{ }x_{2}^{2}+x_{3}^{2}+...+x_{d}^{2}=O(\rho^{-2\alpha}).
\end{equation}
Therefore $G\subset G(+1,a)\cup G(-1,a),$ where $a=\rho-\rho^{-1}$. Using
(76), the relation $\mu(\Pr_{1}(G(+1,a))=O(\rho^{-(d-1)\alpha})$ (see the last
equality in (89)) and taking into account that the expression under the
integral in (76) for $k=1$ is equal to $1+O(\rho^{-\alpha})$ (see (79) and
(89)), we get $\mu(G(+1,a))=O(\rho^{-(d-1)\alpha}).$ Similarly, $\mu
(G(-1,a))=O(\rho^{-(d-1)\alpha}).$ Thus $\mu(G)=O(\rho^{-(d-1)\alpha}).$ Since
$\mid\Gamma_{1}\mid=O(\rho^{d-1}),$ we have
\begin{equation}
\text{ }\mu(\cup_{b\in\Gamma_{1}}(S_{\rho}\cap P_{b})=O(\rho^{-(d-1)\alpha
+d-1})=O(\rho^{-\alpha})\mu(B(\rho)).
\end{equation}

Case 2: $b\notin\Gamma_{1}.$ Then using (88), (85), and $\alpha_{1}=3\alpha,$
we get%
\begin{equation}
\mid x_{1}^{2}-\rho^{2}\mid>2d\rho^{-2\alpha},\text{ }\sum_{k=2}^{d}x_{k}%
^{2}>d\rho^{-2\alpha},\text{ }\max_{k\geq2}\mid x_{k}\mid>\rho^{-\alpha}.
\end{equation}
Therefore $G\subset\cup_{k\geq2}(G(+k,\rho^{-\alpha})\cup G(-k,\rho^{-\alpha
})).$ Now we estimate $\mu(G(+d,\rho^{-\alpha}))$ by using (76). Redenote by
$D$ the set $\Pr_{d}G(+d,\rho^{-\alpha}).$ If $x\in G(+d,\rho^{-\alpha}),$
then according to (85) and (47) the under integral expression in (76) for
$k=d$ is $O(\rho^{1+\alpha}).$ Therefore the first equality in
\begin{equation}
\mu(D)=O(\varepsilon_{1}\mid b\mid^{-1}\rho^{d-2}),\text{ }\mu(G(+d,\rho
^{-\alpha}))=O(\rho^{d-1+\alpha}\varepsilon_{1}\mid b\mid^{-1})
\end{equation}
implies the second equality in (92). To prove the first equality in (92) we
use (77) for $s=d-1$ and $k=1$ and prove the relations $\mu(\Pr_{1}%
D)=O(\rho^{d-2}),$
\begin{equation}
\text{ }\mu(D(x_{2},x_{3},...,x_{d-1}))<6\varepsilon_{1}\mid b\mid^{-1}%
\end{equation}
for $(x_{2},x_{3},...,x_{d-1})\in\Pr_{1}D.$ First relation follows from the
inequalities in (75)). So we need to prove (93). If $x_{1}\in D(x_{2}%
,x_{3},...,x_{d-1})$ then (83) and (84) hold. Subtracting (83) from (84), we
get%
\[
((x_{1}-b_{1})^{2}+x_{2}^{2}+x_{3}^{2}+...+x_{d}^{2})^{l}-((x_{1}^{2}%
+x_{2}^{2}+x_{3}^{2}+...+x_{d}^{2})^{l}%
\]%
\begin{equation}
+F_{k_{1}-1}(x-b)-F_{k_{1}-1}(x)=h,
\end{equation}
where $x_{2},x_{3},...,x_{d-1}$ are fixed . Hence we have two equations (83)
and (94) with respect to two unknown $x_{1}$ and $x_{d}$. Using (47), the
implicit function theorem, and the inequalities $\mid x_{d}\mid>\rho^{-\alpha
},$ $\alpha_{1}>2\alpha$ from (83), we obtain
\begin{equation}
x_{d}=f(x_{1}),\text{ }\frac{df}{dx_{1}}=-\frac{2x_{1}+O(\rho^{-2\alpha
_{1}+\alpha})}{2x_{d}+O(\rho^{-2\alpha_{1}+\alpha})}%
\end{equation}
Substituting $f(x_{1})$ for $x_{d}$ in (94), we get
\[
((x_{1}-b_{1})^{2}+x_{2}^{2}+x_{3}^{2}+...+f^{2}(x_{1}))^{l}-((x_{1}^{2}%
+x_{2}^{2}+x_{3}^{2}+...+f^{2}(x_{1}))^{l}+
\]%
\begin{equation}
F_{k_{1}-1}(x_{1}-b_{1},x_{2},...,x_{d-1},f(x_{1}))-F_{k_{1}-1}(x_{1}%
,...,x_{d-1},f)=h
\end{equation}
Using (47), (95), the first equality in (88), and $\mid x_{d}\mid
>\rho^{-\alpha}$ we see that the derivative (w.r.t. $x_{1}$) of the right-hand
side $a_{l}(x)$ of (96) is%

\begin{equation}
\frac{\partial a_{l}(x)}{\partial x_{1}}=l\mid x-b\mid^{2(l-1)}(2(x_{1}%
-b_{1})+2f(x_{1})f^{^{\prime}}(x_{1}))
\end{equation}

$-l\mid x\mid^{2(l-1)}(2x_{1}+2f(x_{1})f^{^{\prime}}(x_{1}))+O(\rho
^{-2\alpha_{1}+\alpha})(1-\frac{x_{1}+O(\rho^{-2\alpha_{1}+\alpha})}%
{x_{d}+O(\rho^{-2\alpha_{1}+\alpha})}).$

If $l=1,$ then using the first equality in (88) and $\mid x_{d}\mid
>\rho^{-\alpha},$ \ we get%

\begin{equation}
\mid\frac{\partial a_{l}(x)}{\partial x_{1}}\mid>b_{1}=\mid b\mid
\end{equation}
If $l>1,$ then it follows from (83), (84), and (24) that
\[
\mid x\mid^{2l}=\rho^{2l}(1+O(\rho^{-2l-\alpha_{1}})),\text{ }\mid
x-b\mid^{2l}=\rho^{2l}(1+O(\rho^{-2l-\alpha_{1}}))
\]%
\begin{align*}
&  \mid x\mid^{2(l-1)}=\rho^{2(l-1)}(1+O(\rho^{-2l-\alpha_{1}}))^{\frac
{l-1}{l}}=\rho^{2(l-1)}+O(\rho^{-2-\alpha_{1}})\text{ }\\
&  \mid x-b\mid^{2(l-1)}=\rho^{2(l-1)}(1+O(\rho^{-2l-\alpha_{1}}))^{\frac
{l-1}{l}}=\rho^{2(l-1)}+O(\rho^{-2-\alpha_{1}})\text{ }%
\end{align*}
Using these in (97) and arguing as in proof of (98) for $l=1,$ we get the
proof of (98) for $l>1.$ Thus, in any case (98) holds. Therefore from (96), by
using the implicit function theorem, we get $\mid\frac{dx_{1}}{dh}\mid
<\frac{1}{\mid b\mid}.$ This inequality and relation $h\in(-3\varepsilon
_{1},3\varepsilon_{1})$ imply (93). Hence (92) is proved. In the same way we
get the same estimation for $G(+k,\rho^{-\alpha})$ and $G(-k,\rho^{-\alpha})$
for $k\geq2$. Thus

$\mu(S_{\rho}\cap P_{b})=O(\rho^{d-1+\alpha}\varepsilon_{1}\mid b\mid^{-1})$
for $b\notin\Gamma_{1}.$ Since $\varepsilon_{1}=\rho^{-d-2\alpha}$, $\mid
b\mid<2\rho+3$ ( see (88)), using that the number of the vectors of $\Gamma$
satisfying $\mid b\mid<2\rho+3$ is $O(\rho^{d}),$ we get $\mu(\cup
_{b\notin\Gamma_{1}}(S_{\rho}\cap P_{b}))=O(\rho^{2d-1+\alpha}\varepsilon
_{1})=O(\rho^{-\alpha})\mu(B(\rho)).$ Therefore (90) and (72) imply the proof
of (73).

ESTIMATION 3. Here we prove (74). Denote $U_{2\varepsilon}(A_{k,j}(\gamma
_{1,}\gamma_{2},...,\gamma_{k}))$ by $G,$ where $\gamma_{1,}\gamma
_{2},...,\gamma_{k}\in\Gamma(p\rho^{\alpha}),k\leq d-1,$ and $A_{k,j}$ is
defined at the beginning of this section. We turn the coordinate axis so that

$Span\{\gamma_{1,}\gamma_{2},...,\gamma_{k}\}=\{x=(x_{1},x_{2},...,x_{k}%
,0,0,...,0):x_{1},x_{2},...,x_{k}\in\mathbb{R}\}$. Then by (35), we have
$x_{n}=O(\rho^{\alpha_{k}+(k-1)\alpha})$ for $n\leq k,$ $x\in G$. This, (75), and

$\alpha_{k}+(k-1)\alpha<1$ ( see the sixth inequality in (15)) give

$G\subset(\cup_{i>k}(G(+i,\rho d^{-1})\cup G(-i,\rho d^{-1})),$

$\mu(\Pr_{i}(G(+i,\rho d^{-1})))=O(\rho^{k(\alpha_{k}+(k-1)\alpha)+(d-1-k)})$
for $i>k.$ Now using this and (77) for $s=d,$ we prove that
\begin{equation}
\mu(G(+i,\rho d^{-1}))=O(\varepsilon\rho^{k(\alpha_{k}+(k-1)\alpha
)+(d-1-k)}),\forall i>k.
\end{equation}
For this we redenote by $D$ the set $G(+i,\rho d^{-1})$ and prove that
\begin{equation}
\mu((D(x_{1},x_{2},...x_{i-1},x_{i+1},...x_{d}))\leq(42d^{2}+4)\varepsilon
\end{equation}
for $(x_{1},x_{2},...x_{i-1},x_{i+1},...x_{d})\in\Pr_{i}(D)$ and $i>k.$ To
prove (100) it is sufficient to show that if both $x=(x_{1},x_{2}%
,...,x_{i},...x_{d})$ and $x^{^{\prime}}=(x_{1},x_{2},...,x_{i}^{^{\prime}%
},...,x_{d})$ are in $D,$ then $\mid x_{i}-x_{i}^{^{\prime}}\mid\leq
(42d^{2}+4)\varepsilon.$ Assume the converse. Then

$\mid x_{i}-x_{i}^{^{\prime}}\mid>(42d^{2}+4)\varepsilon$. Without loss of
generality it can be assumed that $x_{i}^{^{\prime}}>x_{i}.$ So $x_{i}%
^{^{\prime}}>x_{i}>\rho d^{-1}$ ( see definition of $D$). Since $x$ and
$x^{^{\prime}}$ lie in the $2\varepsilon$ neighborhood of $A_{k,j},$ there
exist points $a,a^{^{\prime}}$ in $A_{k,j}$ such that

$\mid x-a\mid<2\varepsilon$ , $\mid x^{^{\prime}}-a^{^{\prime}}\mid
<2\varepsilon.$ It follows from the definitions of the points $x,$
$x^{^{\prime}},a,$ $a^{^{\prime}}$ that the following inequalities hold:%
\begin{align}
\rho d^{-1}-2\varepsilon &  <a_{i}<a_{i}^{^{\prime}},\text{ }a_{i}^{^{\prime}%
}-a_{i}>42d^{2}\varepsilon,\\
(a_{i}^{^{\prime}})^{2}-(a_{i})^{2}  &  >2(\rho d^{-1}-2\varepsilon
)(a_{i}^{^{\prime}}-a_{i}),\nonumber\\
&  \mid\mid a_{s}\mid-\mid a_{s}^{^{\prime}}\mid\mid<4\varepsilon,\forall
s\neq i.\nonumber
\end{align}
On the other hand the inequalities in (75) hold for the points of $A_{k,j}$ ,
that is, we have $\mid a_{s}\mid<\rho+1,\mid a_{s}^{^{\prime}}\mid<\rho+1.$
Therefore these inequalities and the inequalities in (101) imply $\mid\mid
a_{s}\mid^{2}-\mid a_{s}^{^{\prime}}\mid^{2}\mid<12\rho\varepsilon$ for $s\neq
i$, and hence

$\sum_{s\neq i}\mid\mid a_{s}\mid^{2}-\mid a_{s}^{^{\prime}}\mid^{2}%
\mid<12d\rho\varepsilon<\frac{2}{7}\rho d^{-1}(a_{i}^{^{\prime}}-a_{i}),$%

\begin{equation}
\mid\mid a\mid^{2}-\mid a^{^{\prime}}\mid^{2}\mid>\frac{3}{2}\rho d^{-1}\mid
a_{i}^{^{\prime}}-a_{i}\mid.
\end{equation}
Moreover, using mean value theorem and the relations

$\mid a\mid=\rho+O(1),\mid a^{^{\prime}}\mid=\rho+O(1)$ ( see (75)), we get
\begin{equation}
\mid a\mid^{2l}-\mid a^{^{\prime}}\mid^{2l}=l(\rho+O(1))^{2(l-1)})(\mid
a\mid^{2}-\mid a^{^{\prime}}\mid^{2})
\end{equation}
Let $r_{i}(x)=\lambda_{i}(x)-\mid x\mid^{2l},$ where $\lambda_{i}(x)$ is the
eigenvalues of the matrix $C(x)$ defined in Theorem 2. Hence $r_{1}(x)\leq
r_{2}(x)\leq...\leq$ $r_{b_{k}}(x)$ are the eigenvalues of the matrix
$C^{^{\prime}}(x),$ where $C^{^{\prime}}(x)=C(x)-\mid x\mid^{2l}I$. By
definition of $C^{^{\prime}}(x)$ only diagonal elements of the matrix
$C^{^{\prime}}(x)$ depend on $x$ and they are

$\mid x-d_{i}\mid^{2l}-\mid x\mid^{2l},$ where $d_{i}=h_{i}+t-\gamma-t\in
B_{k}+\Gamma(p_{1}\rho^{\alpha})$. Clearly,%
\begin{equation}
\mid d_{i}\mid<\rho^{\frac{1}{2}\alpha_{d}},\text{ }\mid r_{j}(a^{^{\prime}%
})-r_{j}(a)\mid\leq\parallel C^{^{\prime}}(a^{^{\prime}})-C^{^{\prime}%
}(a)\parallel=\max_{i}\mid a_{i,i}\mid,
\end{equation}
where $C^{^{\prime}}(a^{^{\prime}})-C^{^{\prime}}(a)=(a_{i,j}),$ $a_{i,i}=\mid
a\mid^{2l}-\mid a-d_{i}\mid^{2l}-\mid a^{^{\prime}}\mid^{2l}+\mid a^{^{\prime
}}-d_{i}\mid^{2l},$ and $a_{i,j}=0$ for $i\neq j,$ that is, $C^{^{\prime}%
}(x)-C^{^{\prime}}(x^{^{\prime}})$ is a diagonal matrix. Now we estimate $\mid
a_{i,i}\mid.$ Using mean value theorem and the relations $\mid a\mid
=\rho+O(1),$

$\mid a^{^{\prime}}\mid=\rho+O(1),$ $\mid a-d_{i}\mid=\rho+O(\rho^{\frac{1}%
{2}\alpha_{d}}),\mid a^{^{\prime}}-d_{i}\mid=\rho+O(\rho^{\frac{1}{2}%
\alpha_{d}}),$ we obtain
\[
a_{i,i}=l(\rho+O(\rho^{\frac{1}{2}\alpha_{d}}))^{2(l-1)}(\mid a\mid^{2}-\mid
a^{^{\prime}}\mid^{2})-
\]%
\[
l(\rho+O(\rho^{\frac{1}{2}\alpha_{d}}))^{2(l-1)}(\mid a-d_{i}\mid^{2}-\mid
a^{^{\prime}}-d_{i}\mid^{2})=
\]%
\[
l(\rho+O(\rho^{\frac{1}{2}\alpha_{d}}))^{2(l-1)}(\mid a\mid^{2}-\mid
a-d_{i}\mid^{2}-\mid a^{^{\prime}}\mid^{2}+\mid a^{^{\prime}}-d_{i}\mid^{2})+
\]%
\[
O(\rho^{\frac{1}{2}\alpha_{d}+2l-3})(\mid a\mid^{2}-\mid a^{^{\prime}}\mid
^{2}).
\]
Since $\mid a\mid^{2}-\mid a-d_{i}\mid^{2}-\mid a^{^{\prime}}\mid^{2}+\mid
a^{^{\prime}}-d_{i}\mid^{2}=2(a-a^{^{\prime}},d_{i}),$ we have ( see (104))
$\mid r_{j}(a)-r_{j}(a^{^{\prime}})\mid\leq3l\rho^{\frac{1}{2}\alpha_{d}%
+2l-2}\mid a-a^{^{\prime}}\mid-c_{15}\rho^{\frac{1}{2}\alpha_{d}+2l-3}\mid\mid
a\mid^{2}-\mid a^{^{\prime}}\mid^{2}\mid.$ Therefore using $\frac{1}{2}%
\alpha_{d}<1,$ (103), (102), (101) and $\lambda_{i}(x)=r_{i}(x)+\mid
x\mid^{2l},$ we obtain

$\mid\lambda_{j}(a)-\lambda_{j}(a^{^{\prime}})\mid\geq\mid\mid a\mid^{2l}-\mid
a^{^{\prime}}\mid^{2l}\mid-\mid r_{j}(a)-r_{j}(a^{^{\prime}})\mid>$

$\bigskip$

$l\rho^{2l-1}d^{-1}\mid a_{i}^{^{\prime}}-a_{i}\mid>l\rho^{2l-1}%
42d\varepsilon>6\varepsilon_{1},$ which contradicts the fact that both
$\lambda_{j}(a)$ and $\lambda_{j}(a^{^{\prime}})$ lie in $(\rho^{2}%
-3\varepsilon_{1},\rho^{2}+3\varepsilon_{1})$ ( see the definition of
$A_{k,j}$). Thus (100), hence (99) is proved. In the same way we get the same
formula for $G(-i,\frac{\rho}{d}).$ So $\mu(U_{2\varepsilon}(A_{k,j}%
(\gamma_{1,}\gamma_{2},...,\gamma_{k})))=O(\varepsilon\rho^{k(\alpha
_{k}+(k-1)\alpha)+d-1-k}),$ where

$j=1,2,...,b_{k}(\gamma_{1},\gamma_{2},...,\gamma_{k}),$ and $\gamma
_{1},\gamma_{2},...,\gamma_{k}\in\Gamma(p\rho^{\alpha}).$ From this using that
$b_{k}=O(\rho^{d\alpha+\frac{k}{2}\alpha_{k+1}})$ ( see (43)) and the number
of the vectors $(\gamma_{1},\gamma_{2},...,\gamma_{k})$ for $\gamma_{1}%
,\gamma_{2},...,\gamma_{k}\in\Gamma(p\rho^{\alpha})$ is $O(\rho^{dk\alpha}),$
we obtain (74) if

$d\alpha+\frac{k}{2}\alpha_{k+1}+dk\alpha+k(\alpha_{k}+(k-1)\alpha)+d-1-k\leq
d-1-\alpha$ or%
\begin{equation}
(d+1)\alpha+\frac{k}{2}\alpha_{k+1}+dk\alpha+k(\alpha_{k}+(k-1)\alpha)\leq k
\end{equation}
for $1\leq k\leq d-1$. Dividing both side of (105) by $k\alpha$ and using
$\alpha_{k}=3^{k}\alpha,$ $\alpha=\frac{1}{m},$ $m=3^{d}+d+2$ ( see the end of
the introduction) we see that (105) is equivalent to $\frac{d+1}{k}%
+\frac{3^{k+1}}{2}+3^{k}+k-1\leq3^{d}+2$

The left-hand side of this inequality gets its maximum value at $k=d-1.$
Therefore we need to show that $\frac{d+1}{d-1}+\frac{5}{6}3^{d}+d\leq
3^{d}+4,$ which follows from the obvious inequalities $\frac{d+1}{d-1}%
\leq3,d<\frac{1}{6}3^{d}+1$ for $d\geq2.$

ESTIMATION 4. Here we prove (69). During this estimation we denote by $G$ the
set $S_{\rho}^{^{\prime}}\cap U_{\varepsilon}(Tr(A(\rho))$. Since $V_{\rho
}=S_{\rho}^{^{\prime}}\backslash G$ and (73) holds, it is enough to prove that
$\mu(G)=O(\rho^{-\alpha})\mu(B(\rho)).$ For this we use (75) and prove
$\mu(G(+i,\rho d^{-1}))=O(\rho^{-\alpha})\mu(B(\rho))$ for $i=1,2,...,d$ by
using (76) ( the same estimation for $G(-i,\rho d^{-1})$ can be proved in the
same way). By (47), if $x\in G(+i,\rho d^{-1}),$ then the under integral
expression in (76) for $k=i$ and $a=\rho d^{-1}$ is less than $d+1.$ Therefore
it is sufficient to prove
\begin{equation}
\mu(\Pr(G(+i,\rho d^{-1}))=O(\rho^{-\alpha})\mu(B(\rho))
\end{equation}
Clearly, if $(x_{1},x_{2},...x_{i-1},x_{i+1},...x_{d})\in\Pr_{i}(G(+i,\rho
d^{-1})),$ then

$\mu(U_{\varepsilon}(G)(x_{1},x_{2},...x_{i-1},x_{i+1},...x_{d}))\geq
2\varepsilon$ and by (77), it follows that
\begin{equation}
\mu(U_{\varepsilon}(G))\geq2\varepsilon\mu(\Pr(G(+i,\rho d^{-1})).
\end{equation}
Hence to prove (106) we need to estimate $\mu(U_{\varepsilon}(G)).$ For this
we prove that
\begin{equation}
U_{\varepsilon}(G)\subset U_{\varepsilon}(S_{\rho}^{^{\prime}}),U_{\varepsilon
}(G)\subset U_{2\varepsilon}(Tr(A(\rho))),U_{\varepsilon}(G)\subset
Tr(U_{2\varepsilon}(A(\rho))).
\end{equation}
The first and second inclusions follow from $G\subset S_{\rho}^{^{\prime}}$
and $G\subset U_{\varepsilon}(Tr(A(\rho)))$ respectively (see definition of
$G$ ). Now we prove the third inclusion in (108). If $x\in U_{\varepsilon
}(G),$ then by the second inclusion of (108) there exists $b$ such that $b\in
Tr(A(\rho)),$ $\mid x-b\mid<2\varepsilon.$ Then by the definition of
$Tr(A(\rho))$ there are $\gamma\in\Gamma$ and $c\in A(\rho)$ such that
$b=\gamma+c$. Therefore $\mid x-\gamma-c\mid=\mid x-b\mid<2\varepsilon,$

$x-\gamma\in U_{2\varepsilon}(c)\subset U_{2\varepsilon}(A(\rho)).$ This
together with $x\in U_{\varepsilon}(G)\subset U_{\varepsilon}(S_{\rho
}^{^{\prime}})$ (see the first inclusion of (108)) give $x\in
Tr(U_{2\varepsilon}(A(\rho)))$ , i.e., the third inclusion in (108) is proved.
The third inclusion, Lemma 2(c), and (74) imply that

$\mu(U_{\varepsilon}(G))=O(\rho^{-\alpha})\mu(B(\rho))\varepsilon.$ This and
(107) imply the proof of (106)$\diamondsuit$

\end{document}